\documentclass[a4paper,fleqn,sort&compress]{cas-sc}

\usepackage{mathtools}
\usepackage{esint}
\usepackage{graphicx}
\usepackage{graphics}
\usepackage{multirow}
\usepackage{bm}
\usepackage{listings}
\usepackage{color}
\usepackage[colorlinks]{hyperref}
\usepackage{cancel}
\usepackage{braket}
\usepackage{relsize}
\usepackage{bm}
\usepackage{amsmath}
\usepackage{amsbsy}
\usepackage{sidecap}
\usepackage{subcaption}

\def \bR{\mathbf{R}}
\def \bL{\mathbf{L}}
\def \Bz{\mathfrak{B}}
\def \bx{\mathbf{x}}
\def \bk{\mathbf{k}}
\def \bg{\mathbf{g}}
\def \cI{\mathcal{I}}
\def \bE{\mathbf{E}}

\def \bQ{\mathbf{Q}}

\usepackage{placeins}

\usepackage[numbers]{natbib}

\ExplSyntaxOn
\cs_set:Npn \__first_footerline:
{
  \group_begin:
  \small
  \sffamily
  \ifnum\theblind>0\relax
  \else
    \__short_authors: :~
  \fi
  { \rmfamily \itshape Preprint~ submitted ~to ~arXiv }
  \group_end:
}
\ExplSyntaxOff

\begin{document}
\let\WriteBookmarks\relax
\def\floatpagepagefraction{1}
\def\textpagefraction{.001}
\shorttitle{Unified open-boundary electrostatics in real-space DFT}
\shortauthors{Rajat et~al.}

\title [mode = title]{Unified open-boundary electrostatics in real-space density functional theory}

\author[1]{Rajat Kumar}[]

\affiliation[1]{organization={Department of Civil Engineering, Indian Institute of Technology Roorkee},
                city={Roorkee},
                postcode={247667}, 
                state={Uttarakhand},
                country={India}}

\author[2,3]{David Codony}[]

\affiliation[2]{organization={Laboratori de Càlcul Numèric (LaCàN), Universitat Politècnica de Catalunya},
                city={Barcelona},
                postcode={E-08034},
                country={Spain}}
                
\affiliation[3]{organization={Institut de Matemàtiques de la UPC -BarcelonaTech (IMTech)},
                city={Barcelona},
                postcode={08034},
                country={Spain}}                             

\author[4,5]{Phanish Suryanarayana}[]

\affiliation[4]{organization={College of Engineering, Georgia Institute of Technology
},
                city={Atlanta},
                postcode={30332}, 
                state={Georgia},
                country={USA}}
                
\affiliation[5]{organization={College of Computing, Georgia Institute of Technology
}, 
                city={Atlanta},
                postcode={30332}, 
                state={Georgia},
                country={USA}}

\author[1]{Abhiraj Sharma}[orcid=0009-0001-7249-6864]
             
\cormark[1]                
\ead{abhiraj.sharma@ce.iitr.ac.in}

\cortext[cor1]{Corresponding author}

\begin{abstract}
We present an electrostatic formulation in real-space density functional theory that provides a systematic and unified treatment of the open-boundary electrostatics of isolated and partially periodic systems, including in the presence of an applied uniform electric field along the open (finite) directions. Specifically, we formulate a local electrostatic energy functional whose stationarity yields the Poisson equation for the electrostatic potential, subject to periodic and Dirichlet boundary conditions along the periodic and open directions, respectively. Using a Green's function approach, we derive analytical expressions for the Dirichlet values arising from the total charge density of the system. We also derive the expressions for the energy, atomic forces, and stress tensor. We implement the resulting expressions within the large-scale parallel real-space SPARC electronic structure code. Using representative examples, we verify the accuracy and efficiency of the framework, demonstrating exponential convergence of the computed quantities with vacuum size and excellent agreement with established plane-wave codes while requiring significantly less vacuum at comparable accuracy. Since no existing implementation provides the stresses for such systems, we instead verify them against numerical derivatives of the energy, finding close agreement. Finally, we apply the framework to compute static polarizabilities and piezoelectric coefficients, obtaining very good agreement with values reported in the literature.
\end{abstract}

\begin{keywords}
Real-space density functional theory \\
Electrostatics \\
Electric field \\
Poisson equation \\
Dirichlet boundary conditions
\end{keywords}

\maketitle
\section{Introduction}
Kohn--Sham density functional theory (DFT)~\cite{hohenberg1964inhomogeneous,kohn1965self} has become the workhorse of materials and chemical sciences research owing to its predictive power and high accuracy-to-cost ratio relative to other electronic structure methods \cite{burke2012dft,becke2014dft}. Within the Kohn--Sham formulation, the electrostatics poses a unique challenge owing to the long-range nature of the $1/r$ Coulomb kernel and the associated singularity at $r \rightarrow 0$. The former causes the potentials and interaction energies to diverge individually  owing to the infinite lattice summations along the periodic directions, renders the evaluation complexity of the interaction energy quadratic in system size, and often necessitates large vacuum regions along the open directions of isolated (e.g., molecules) and partially periodic (e.g., 1D nanowires and 2D slabs) systems; the latter results in a divergent potential at the nuclear sites and infinite self-energies of the point nuclei. The singularity and its associated difficulties are largely circumvented within the pseudopotential approximation \cite{pickett1989pseudopotential}, wherein the point nuclei are replaced by smooth ionic cores. The challenges arising from the long-range nature of the Coulomb kernel, however, are addressed differently across discretization techniques for solving the Kohn--Sham equations.

Among the most widely adopted discretization techniques is the plane-wave pseudopotential method \cite{Martin2004,kresse1996efficient,clark2005first,gonze2002first,giannozzi2009quantum,ismail2000new,gygi2008architecture,valiev2010nwchem}, which offers the orthonormality and completeness of the Fourier basis, systematic convergence, and access to the fast Fourier transform (FFT). Within this framework, the ionic and electronic contributions to the electrostatics are handled separately. The ion--ion interactions are evaluated using the Ewald lattice sum \cite{ewald1921berechnung,ihm1979momentum,pickett1989pseudopotential}. The electron--electron interaction is obtained by solving the corresponding Poisson equation in reciprocal space using FFTs, while the electron--ion interaction enters as an external potential assembled in reciprocal space. The divergent zero-wavevector components of the three contributions are individually removed, with the residual constants cancelling exactly for a charge-neutral system. Overall, the cost of the electrostatic evaluation is dominated by the FFTs, thereby scaling as $\mathcal{O}(N \log N)$ with the number of atoms $N$. 

However, since the plane-wave method assumes periodicity in all directions, isolated and partially periodic systems require the introduction of artificial periodicity with vacuum regions along the open directions. The spurious interactions between artificial images that this introduces are governed by the leading non-vanishing multipole moment of the charge density, and decay only algebraically with the supercell dimension \cite{makov1995periodic}. A separate difficulty arises when a uniform electric field is applied along the open directions, since the associated linearly varying potential is incompatible with periodicity altogether. A variety of correction schemes have been developed to mitigate these issues. For 2D slabs with a net out-of-plane dipole moment and/or subjected to a uniform electric field along the surface normal, an artificial sawtooth potential is inserted in the vacuum region \cite{kunc1983external,neugebauer1992adsorbate,bengtsson1999dipole,meyer2001ab}. For 1D nanowires and 2D slabs, corrections based on truncation of the Coulomb kernel in reciprocal space \cite{rozzi2006exact,ismail2006truncation} and on the difference between the periodic and exact potentials \cite{dabo2008electrostatics} have been devised, some of which extend to isolated systems as well \cite{rozzi2006exact,dabo2008electrostatics}. For isolated systems specifically, modified reciprocal-space kernels constructed from a screening function \cite{martyna1999reciprocal} and real-space convolution on a doubled grid \cite{barnett1993born} have been developed. These schemes, however, still require substantial vacuum padding and complicate the formulation, so that each response property must be reformulated separately---as in the extensions of density functional perturbation theory to partially periodic systems \cite{sohier2017density,rivano2024density}. The vacuum requirement is particularly severe for the unoccupied states entering many-body perturbation theories such as GW and the RPA, whose quasiparticle energies converge far more slowly with vacuum size \cite{freysoldt2008screening}. Besides these limitations, the non-local nature of the Fourier basis necessitates global communication in the FFTs, limiting parallel scalability on high-performance computing platforms.  

The above limitations of the plane-wave method have motivated the development of alternative solution strategies based on localized representations \cite{becke1989basis,white1989finite,chel1994fdpp,seitsonen1995real,tsuchida1995electronic,briggs1996real,fattebert1999finite,arias1999wav,shimojo2001linear,skylaris2005introducing,pask2005femeth,bowler2006recent,castro2006octopus,genovese2008daubechies,iwata2010massively,suryanarayana2010non,suryanarayana2011mesh,lin2012adaptive,Ghosh2017cluster,Ghosh2017extended,xu2018discrete,motamarri2020dft}, among which the real-space finite-difference method \cite{beck2000rsmeth,saad2010esmeth} is arguably the most widely adopted technique. As a basis-free approach, it maximizes computational locality by discretizing all spatial quantities on a uniform grid. Moreover, the localized representation facilitates linear-scaling algorithms \cite{pratapa2015spectral,suryanarayana2018sqdft} and massive parallelization \cite{gavini2023roadmap} on modern computing platforms. The method naturally accommodates both periodic and Dirichlet boundary conditions, allowing for the accurate and efficient treatment of periodic, partially periodic, and isolated systems, including those with non-traditional symmetries \cite{banerjee2016cyclic,ghosh2019symmetry,sharma2021real}. Within this framework, the non-local Coulomb kernel in the electrostatics is replaced by a differential operator, the ionic charges are represented by localized pseudocharge densities \cite{pask2005femeth,Gavini2007,suryanarayana2010non}, and the electrostatic potential for the total charge density is obtained from the solution of the Poisson equation \cite{Ghosh2017cluster,Ghosh2017extended}. In some real-space electrostatic formulations \cite{alemany2004real}, however, only the Hartree potential is obtained using the Poisson equation, with the divergent electron--ion and ion--ion contributions treated in reciprocal space---via an FFT of the local pseudopotential and the Ewald method, respectively. In either case, the cost of the electrostatic evaluation is dominated by the solution of the Poisson equation, which scales as $\mathcal{O}(N)$ with iterative or multigrid methods.

This solve, however, requires Dirichlet values to be prescribed for isolated and partially periodic systems. These are the values taken by the electrostatic potential on the open boundaries, and are often non-vanishing due to the presence of multipole moments and/or an applied electric field along the open directions. The analytical forms of these boundary conditions have been developed for isolated systems using multipole expansion in spherical coordinates \cite{hirose2005first}; 1D nanowires using multipole expansion in cylindrical coordinates \cite{han2008real}; and 2D slabs using the asymptotic behavior of the electrostatic potential \cite{natan2008real}. These works, however, lack a unifying formulation across dimensionalities, and derive the Dirichlet values only for the Hartree potential, treating the electron--ion and ion--ion interactions through auxiliary constructs borrowed from the periodic setting. Moreover, none of these approaches incorporates an external electric field along the open directions. A recent real-space formulation \cite{ramakrishnan2025real} accounts for a uniform out-of-plane electric field in 2D slabs, but neither incorporates multipole effects nor extends to other dimensionalities. Furthermore, to the best of our knowledge, a formulation of the stress tensor for systems with a non-vanishing electrostatic potential on the open boundaries has not been developed within either the plane-wave or the real-space framework.

These limitations motivate the present work, in which we develop an electrostatic formulation in real-space density functional theory that provides a systematic and unified treatment of the open-boundary electrostatics of isolated and partially periodic systems, including in the presence of an applied uniform electric field along the open directions. In particular, we formulate an electrostatic energy functional whose stationarity yields the Poisson equation for the electrostatic potential---arising from the total charge density as well as the applied uniform electric field---subject to periodic and Dirichlet boundary conditions along the periodic and open directions, respectively. The Dirichlet values are prescribed from the analytical form of the electrostatic potential in the region where the charge density vanishes, with the corresponding expressions derived using a Green's function approach. Using the derived electrostatics, we develop the real-space formulation of Kohn--Sham DFT, deriving the electronic ground-state equations along with the expressions for the ground-state energy, atomic forces, and stress tensor. We implement the resulting boundary conditions and the modified expressions for the ground-state quantities in SPARC \cite{xu2021sparc,zhang2024sparc}, a large-scale parallel real-space electronic structure code. The accuracy and efficiency of the formulation and its implementation are verified using representative systems, and the framework is then applied to compute their static polarizabilities and piezoelectric coefficients.
 
The remainder of this manuscript is organized as follows. In Sec.~\ref{Sec:Electrostatics}, we derive the local variational formulation of electrostatics, on which the real-space formulation of DFT presented in Sec.~\ref{Sec:DFT} is based. Implementation details are provided in Sec.~\ref{Sec:implementation}, and the accuracy and efficiency of the framework are verified in Sec.~\ref{Sec:numericaltests}. In Sec.~\ref{Sec:applications}, we compute the static polarizabilities and piezoelectric coefficients of the systems considered. Finally, we provide concluding remarks in Sec.~\ref{Sec:conclusion}.
\section{Electrostatic formulation} \label{Sec:Electrostatics}
\begin{figure}[!htbp]
        \centering
        \includegraphics[width=0.3\textwidth,keepaspectratio=true]{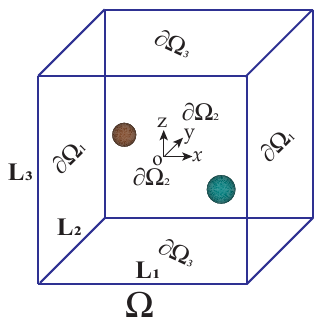}  
    \caption{Illustration of a $2$-atom unit cell $\Omega$ centered at the origin and defined by the vectors $\bL_1$, $\bL_2$, and $\bL_3$. The pair of boundary faces of $\Omega$ spanned by $\{\bL_\beta:\beta \neq \alpha \}$ is denoted by $\partial \Omega_\alpha$, $\alpha,\beta \in \{1,2,3\}$.}  \label{Fig:unit_cell} 
\end{figure}
Consider a $d$-dimensional (where $d \in \{0,1,2,3\}$) charge-neutral system spanned by the vectors $\{\bL_\alpha : \alpha \in \cI_p \subseteq \{1,2,3\}, |\cI_p|=d \}$ in the periodic directions and defined by the vectors $\{\bL_\alpha : \alpha \in \cI_o = \{1,2,3\}\setminus \cI_p, |\cI_o|=3-d\}$ in the open directions. Let $\Omega$ be the $N$-atom unit cell of the system, and let $\partial\Omega_\alpha$ denote the pair of boundary faces of $\Omega$ spanned by $\{\bL_\beta : \beta \in \{1,2,3\}, \beta \neq \alpha\}$, as shown in Fig.~\ref{Fig:unit_cell}. For all $\alpha \in \cI_o$, the vectors $\bL_\alpha$ are assumed long enough that the electron density vanishes on and beyond $\partial \Omega_\alpha$. Let the system be subjected to a uniform electric field $\bE$ applied along the open directions (i.e., $\bE \cdot \bL_\alpha = 0 \;\; \forall \, \alpha \in \cI_p$), with the source of the field placed sufficiently distant to remain electronically disconnected from the system. For this prototypical system, we now develop a local variational formulation of electrostatics in real space incorporating the non-vanishing electrostatic potential on $\partial \Omega_\alpha$, $\alpha \in \cI_o$, through surface integrals and Dirichlet boundary conditions derived from the analytical form of the potential in the region where the charge density vanishes. 
\subsection{Electrostatic energy}
Within the formalism of pseudopotential Kohn--Sham DFT \cite{Martin2004,kohn1965self}, the electrostatic energy functional associated with $\Omega$ can be written as:
\begin{align}
\mathcal{E}_\text{el}[\rho;\bR,\bE] =& \frac{1}{2} \int_{\Omega} \mathrm{d} \bx \, \rho(\bx)\int_{\mathbb{R}^3} \mathrm{d} \bx' \, \frac{ \rho(\bx')}{|\bx-\bx'|}  + \sum_{I=1}^{M} \int_{\Omega} \mathrm{d} \bx \, \rho(\bx) V_I(\bx,\bR_I) + \frac{1}{2} \sum_{I=1}^{N} \sum_{\substack{J=1\\J\neq I}}^{M} \frac{Z_I Z_J}{|\bR_I-\bR_J|} \nonumber \\
&+ \int_\Omega \mathrm{d} \bx \, \left(\rho(\bx)+ \sum_{I=1}^M Z_I \delta(\bx-\bR_I) \right) \bx \cdot \bE \,, \label{Eqn:KSelecenergy}
\end{align}
where the first three terms are the electron--electron, electron--ion, and ion--ion interaction energies, respectively; and the last term is the interaction energy of the total (electron+ion) charge density with the applied uniform electric field $\bE$. In the expression, $\rho$ is the charge density of the valence electrons, $\bR$ is the set of positions of $M$ nuclei in the full system, $\delta$ is the Dirac delta distribution, and $V_I$ is the local component of the pseudopotential associated with the $I^\text{th}$ ion which is centered at $\bR_I$ and has a net charge $Z_I$. The factor of $1/2$ in the first and third terms corrects for the double counting of pairwise interactions. Note that a positive sign is adopted for the potential term arising from $\bE$, consistent with the convention that the electronic charge is taken as positive in DFT.

Each of the first three terms in Eq.~\eqref{Eqn:KSelecenergy} diverges for $d = 1,2,3$ owing to the long-range Coulomb kernel; their sum, however, is convergent since the system is charge-neutral. Following Refs.~\cite{pask2005femeth,suryanarayana2010non,Ghosh2017extended}, we exploit this property, together with the smooth, non-singular nature of the pseudopotentials, to recast the electrostatic energy functional as
\begin{align}
\mathcal{E}_\text{el}[\rho;\bR,\bE] =& \frac{1}{2} \int_\Omega \mathrm{d} \bx \, (\rho(\bx)+b(\bx,\bR)) \int_{\mathbb{R}^3} \mathrm{d} \bx' \, \frac{\rho(\bx')+b(\bx',\bR)}{|\bx-\bx'|} + \mathcal{E}_\text{sc}(\bR) + \int_\Omega \mathrm{d} \bx \, (\rho(\bx) + b(\bx,\bR) ) \bx \cdot \bE \,, \label{Eqn:Elecenergyrefor}
\end{align}
where $b = \sum_{I=1}^M b_I$ is the total ionic pseudocharge density, with $b_I = -\frac{1}{4 \pi} \nabla^2 V_I$ the spherically symmetric and localized ionic pseudocharge density of the $I^\text{th}$ ion. In the above expression, the first term is the electrostatic energy of the total charge density $\rho+b$, and the last term is the energy of its interaction with the applied electric field. The remaining term, $\mathcal{E}_\text{sc}$, is the self-interaction and overlap-correction energy, which removes the self-interaction energy of each ion and corrects the ion-ion interaction energy in regions where the pseudocharge densities of different ions overlap.

Although the divergence issue is resolved in Eq.~\eqref{Eqn:Elecenergyrefor}, the non-local nature of the Coulomb kernel in the first term causes its evaluation in real space to scale quadratically with the number of atoms. Moreover, the functional, as written, does not admit a local variational formulation---a structure that, among other advantages, replaces the non-local Coulomb kernel with a local differential operator whose inversion reproduces the same long-range interaction exactly, while its stationarity ensures consistency between the energy and its derivatives. Both limitations can be overcome by recasting the electrostatic energy functional as the extremum of a local functional, as described below.
\subsection{Local reformulation}
\begin{figure}[!htbp]
	\centering
	\includegraphics[width=0.8\textwidth,keepaspectratio=true]{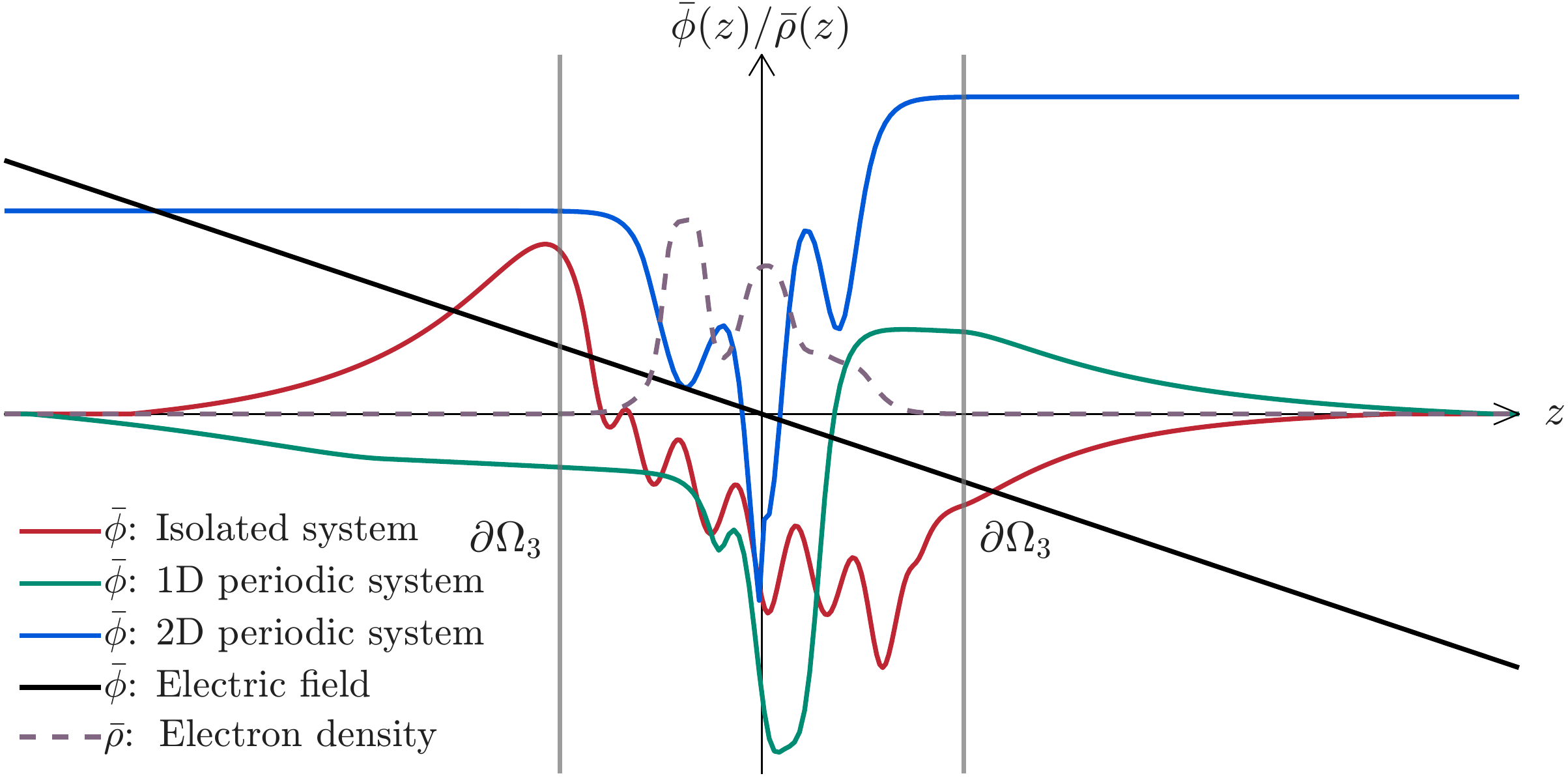}
    \caption{Schematic representation of the electrostatic potential $\bar{\phi}(z)$, averaged over the plane perpendicular to the open $z$-direction, for isolated (red), 1D periodic (green), and 2D periodic (blue) systems. The black line represents the linear potential due to an applied uniform electric field, and the two vertical gray lines mark the boundary planes ($\partial \Omega_3$) outside which the average electron density $\bar{\rho}(z)$ (dashed curve) vanishes.} \label{Fig:phi_illustration}
\end{figure}
We now reformulate the electrostatic energy functional as a variational problem in terms of an auxiliary function $\phi$ that reduces to the electrostatic potential at stationarity. Unlike previous local variational electrostatic formulations \cite{Gavini2007,suryanarayana2010non,Ghosh2017cluster,Ghosh2017extended,ramakrishnan2025real}, which assume that $\phi$ and/or $\nabla \phi$ vanish on the boundary, we do not invoke such an assumption. This is because $\phi$ and/or $\nabla \phi$ can decay slowly or not at all along the open directions---as occurs, for instance, for systems with appreciable multipole moments along the open directions and/or under the applied electric field $\bE$---and therefore extend well beyond the support of the charge density, as illustrated in Fig.~\ref{Fig:phi_illustration}. We accordingly recast the electrostatic energy functional as the following local variational problem, which accommodates any prescribed values of $\phi$ on $\partial \Omega_\alpha$, $\alpha \in \cI_o$: 
\begin{align}
\mathcal{E}_{\text{el}}[\rho;\bR,\bE] = \max_{\phi} \bigg\{&-\frac{1}{8 \pi} \int_{\Omega} \mathrm{d} \bx \, |\nabla \phi(\bx)|^2 + \frac{1}{8\pi} \sum_{\alpha \in \cI_o}\int_{\partial \Omega_\alpha} \mathrm{d}s(\bx) \, \big((\phi(\bx) - \bx \cdot \bE) \nabla \phi(\bx) + \phi(\bx) \, \bE \big) \cdot \hat{\mathbf{n}} \nonumber \\
&+\int_\Omega \mathrm{d} \bx \, (\rho(\bx)+b(\bx,\bR)) \phi(\bx) \bigg\} + \mathcal{E}_{\text{sc}}(\bR)\,,
\end{align}
where $\hat{\mathbf{n}}$ is the outward unit normal to $\partial \Omega_\alpha$, and the maximization is over $\phi \in \mathcal{A} \subset H^1(\Omega)$. Here, $H^1(\Omega)$ denotes the Sobolev space of square-integrable functions on $\Omega$ with square-integrable weak first derivatives, and $\mathcal{A}$ is the affine space of functions that are periodic in the directions $\alpha \in \cI_p$ and satisfy the Dirichlet conditions $\phi(\bx) = \phi_0(\bx) + \bx \cdot \bE$ on $\partial\Omega_\alpha$, $\alpha \in \cI_o$. The function $\phi_0$ is the electrostatic potential due to the total charge density, harmonic on and beyond $\partial\Omega_\alpha$, $\alpha \in \cI_o$. The stationary point $\phi$ of the above functional is the electrostatic potential due to the total charge density of the system and the applied electric field $\bE$; the corresponding stationarity condition is the Poisson equation:
\begin{align}
-\frac{1}{4 \pi} \nabla^2 \phi(\bx) = \rho(\bx)+b(\bx,\bR)\,, \quad \bx \in \Omega \,. \label{Eqn:Poisson}
\end{align}
Finally, substituting this $\phi$ back into the functional yields the closed-form electrostatic energy:
\begin{align}
\mathcal{E}_{\text{el}}[\rho;\bR,\bE] = \frac{1}{2} \int_\Omega \mathrm{d} \bx \, (\rho(\bx)+b(\bx,\bR)) (\phi(\bx)+\bx \cdot \bE) + \mathcal{E}_{\text{sc}}(\bR) \,. \label{Eqn:Electrostaticenergyclosedform}
\end{align}

Next, we derive the expression for $\phi_0$ for the isolated and partially periodic systems, valid on and beyond $\partial \Omega_\alpha, \alpha \in \cI_o$. In order to do so, we first write the solution of Poisson equation in terms of the Green's function:
\begin{align}
\phi(\bx) = -4 \pi \int_{\Omega} \mathrm{d} \bx' \, \left(\rho(\bx')+b(\bx',\bR) \right) G(\bx,\bx') + \bx \cdot \bE\,, \label{Eq:PoissonSolGreens}
\end{align}
where $G$ is the symmetric Green's function, periodic along the directions $\alpha \in \cI_p$, satisfying
\begin{align}
\nabla^2 G(\bx,\bx') = \frac{1}{|\Omega|} \prod_{\beta \in \cI_o}\delta(x_\beta-x'_\beta) \prod_{\alpha \in \cI_p} \sum_{m \in \mathbb{Z}} e^{i m \bQ_\alpha \cdot (\bx-\bx')} \,, \quad \bx,\bx' \in \Omega \,. \label{Eqn:Greensgoverningeqn}
\end{align}
Here, $\phi_0$ corresponds to the the first term of Eq.~\eqref{Eq:PoissonSolGreens} evaluated on and beyond $\partial \Omega_\alpha$, $\alpha \in \cI_o$,  $\bQ_\alpha$ is the reciprocal lattice vector satisfying $\bQ_\alpha \cdot \bL_\beta = 2 \pi \delta_{\alpha \beta}$, and $|\Omega|$ denotes the periodic area, periodic length, and unity for 2D periodic, 1D periodic, and isolated systems, respectively. In deriving \eqref{Eqn:Greensgoverningeqn}, we express the Dirac delta along the periodic directions in terms of plane waves. Now using the fact that plane waves are eigenfunctions of the Laplacian operator under periodic boundary conditions, we construct a Green's function ansatz separately for each type of system and use it to derive the respective $\phi_0$, as discussed below.
\paragraph{0D isolated system.} Since an isolated system (e.g., a molecule) is finite and non-periodic in all three directions, its Green's function is the free-space fundamental solution of the Poisson equation in $\mathbb{R}^3$ given by:
\begin{align}
G^{(0D)}(\bx,\bx') = -\frac{1}{4 \pi |\bx-\bx'|}\,.
\end{align}
Expanding the Coulomb kernel in the above expression in terms of spherical multipole moments and using the fact that the total charge density has a compact support in $\Omega$, we get the expression for $\phi_0$ for the isolated system as:
\begin{align}
\phi_0^{(0D)}(\bx)=\sum_{\ell=1}^{\ell_{\text{max}}} \sum_{m=-\ell}^{\ell} \frac{4 \pi}{(2\ell+1) \mathrm{x}^{\ell+1}}Y_{\ell m} \left ( \frac{\bx}{\mathrm{x}} \right ) \int_{\Omega} \mathrm{d} \bx' \, \left(\rho(\bx')+b(\bx',\bR)\right) {\mathrm{x}'}^{\ell} Y_{\ell m}^* \left ( \frac{\bx'}{\mathrm{x}'} \right ) \,, \, \, \, \bx \in \partial \Omega_\beta, \beta \in \cI_o \,,
\end{align}
where $\mathrm{x}$ is the distance of $\bx$ from the origin, $Y_{\ell m}$ is the spherical harmonic function, and $\ell_\text{max} \in \mathbb{N}$ is a truncation parameter. The series converges provided the evaluation points $\bx$ lie outside a sphere enclosing the support of the total charge density, a condition met in practice since the vectors $\{\bL_\alpha : \alpha \in \cI_o\}$ are chosen such that the charge density decays well within the domain. The $\ell=0$ term is omitted in the above expression, as it is proportional to the total charge and hence vanishes for a charge-neutral system. The decay of the potential is governed by the lowest non-vanishing multipole moment; for a system with a non-vanishing dipole moment this gives $\mathcal{O}(\mathrm{x}^{-2})$, as illustrated in Fig.~\ref{Fig:phi_illustration}.
\paragraph{1D periodic system.} Since a 1D periodic system (e.g., a nanowire) is periodic in one direction and open in the remaining ones, its Green's function ansatz can be taken as a linear combination of 1D plane waves as shown below:
\begin{align}
G^{(1D)}(\bx,\bx') = \sum_{n \in \mathbb{Z}} c_n(\bx_{2D},\bx'_{2D}) \, e^{i  n Q_\alpha (x_\alpha-x'_\alpha)} \,, \quad \alpha \in \cI_p \,,
\end{align}
where $\bx_{2D} = (x_{\beta_1}, x_{\beta_2})$ is a 2D position vector with $\beta_1, \beta_2 \in \cI_o$, $x_\alpha$ is the $\alpha$th component of $\bx$, and $Q_\alpha = \frac{2 \pi}{L_\alpha}$, $L_\alpha$ being the length of the lattice vector $\bL_\alpha$. Substituting the above expression in Eq.~\eqref{Eqn:Greensgoverningeqn}, we obtain the following governing equation for the coefficients $c_n$:
\begin{align}
\left[\frac{\partial^2}{\partial x_{\beta_1}^2} + \frac{\partial^2}{\partial x_{\beta_2}^2} - n^2 Q_\alpha^2 \right] c_n(\bx_{2D},\bx'_{2D}) = \frac{\delta(\bx_{2D} - \bx'_{2D})}{L_\alpha} \,, \quad \forall \, n \in \mathbb{Z} \,.
\end{align}
Therefore, $L_\alpha c_n$ is the Green's function of the screened Poisson equation in $2$D with the screening length $\frac{1}{n Q_\alpha}$. Solving the above differential equation gives:
\begin{align}
c_n(\bx_{2D},\bx'_{2D}) = 
\begin{cases}
\frac{1}{2 \pi L_\alpha} \log|\bx_{2D}-\bx'_{2D}| \,,& n = 0 \\
-\frac{1}{2 \pi L_\alpha} K_0\left(|n| Q_\alpha |\bx_{2D}-\bx'_{2D}|\right)\,,& n \in \mathbb{Z}\setminus \{0\}
\end{cases} \,,
\end{align}
where $K_0$ is the zeroth-order modified Bessel function of the second kind; this solution was given for a periodic line of charges by Lennard-Jones and Dent \cite{lennard1928cohesion}. Now expanding the logarithm term using cylindrical multipole moments and utilizing the facts that the Green's function has the same periodicity as the system and that the system is charge neutral, we obtain the expression for $\phi_0$ for the 1D periodic system as:
\begin{align}
\phi_0^{(1D)}(\bx) =& \frac{2}{L_\alpha} \Bigg[\sum_{m=1}^{m_{\text{max}}}\Re \Bigg\{\frac{e^{-im \theta}}{m {\mathrm{x}}_{2D}^{m}} \int_\Omega \mathrm{d} \bx' \, \left(\rho(\bx')+b(\bx',\bR)\right) \mathrm{x'}_{\!\!\! 2D}^{m} e^{im {\theta}'} \Bigg\} + \sum_{n=1}^{n_{\text{max}}}\Re \Bigg\{2 e^{i n Q_\alpha x_\alpha} \int_\Omega \mathrm{d} \bx' \, e^{-i n Q_\alpha x'_\alpha} \nonumber \\
&\times \left(\rho(\bx')+b(\bx',\bR)\right) K_0\left(n Q_\alpha |\bx_{2D}-\bx'_{2D}|\right) \Bigg\}\Bigg] \,, \quad \alpha \in \cI_p\,, \bx \in \partial \Omega_\beta, \beta \in \cI_o \,,
\end{align}
where $\theta = \tan^{-1}(x_{\beta_2}/x_{\beta_1})$, ${\mathrm{x}}_{2D}$ is the magnitude of $\bx_{2D}$, and $m_\text{max} \in \mathbb{N}$ and $n_\text{max} \in \mathbb{N}$ are the truncation parameters. The first series is convergent provided the evaluation points $\bx$ lie outside a cylinder enclosing the support of the total charge density, a condition met in practice since the vectors $\{\bL_\alpha : \alpha \in \cI_o\}$ are chosen such that the charge density decays well within the domain. The second series is also convergent because $K_0$ decays exponentially with increasing $n$. It is clear from the above expression that the electrostatic potential of a neutral 1D periodic system decays as $\mathcal{O}(\mathrm{x}_{2D}^{-m_0})$ along the open directions, where $m_0$ is the order of the lowest non-vanishing cylindrical multipole moment; for a system with a non-vanishing cylindrical dipole moment, this gives $\mathcal{O}(\mathrm{x}_{2D}^{-1})$, as illustrated in Fig.~\ref{Fig:phi_illustration}. It is important to note that the $K_0$ function can be separated into products of single-argument modified Bessel functions via Graf's addition theorem \cite{abramowitz1964handbook}, as done in previous work \cite{han2008real}. We refrain from doing so, however, as it introduces an additional expansion index and the resulting series typically converges slowly.
\paragraph{2D periodic system.} Since a 2D periodic system (e.g., a slab) is periodic in two directions and open in the remaining one, its Green's function ansatz can be taken as a linear combination of 2D plane waves as shown below:
\begin{align}
G^{(2D)}(\bx,\bx') = \sum_{m,n \in \mathbb{Z}} c_{mn}(x_\beta,x'_\beta) \, e^{i (m {\bQ}_{\alpha_1}+n {\bQ}_{\alpha_2}) \cdot (\bx-\bx')}\,, \quad \beta \in \cI_o \,, \, \alpha_1,\alpha_2 \in \cI_p \,,
\end{align}
where $\bQ_{\alpha_1} = 2\pi \frac{\bL_{\alpha_2} \times \bL_{\beta}}{\bL_{\alpha_1} \cdot (\bL_{\alpha_2} \times \bL_{\beta})}$ and $\bQ_{\alpha_2} = 2\pi \frac{\bL_{\alpha_1} \times \bL_{\beta}}{\bL_{\alpha_2} \cdot (\bL_{\alpha_1} \times \bL_{\beta})}$. Substituting the above expression in Eq.~\eqref{Eqn:Greensgoverningeqn}, we obtain the following governing equation for the coefficients $c_{mn}$:
\begin{align}
\left[\frac{\partial^2}{\partial x_\beta^2} - Q_{mn}^2 \right] c_{mn}(x_\beta,x_\beta') = \frac{\delta(x_\beta-x_\beta')}{A} \,, \quad  \forall \,  m,n \in \mathbb{Z} \,,
\end{align}
where $A = |\bL_{\alpha_1} \times \bL_{\alpha_2}|$ is the cross-sectional area of $\partial \Omega_\beta, \beta \in \cI_o$ and  $Q_{mn}$ is the length of $\bQ_{mn}=m \bQ_{\alpha_1} + n \bQ_{\alpha_2}$. Therefore, $A c_{mn}$ is the Green's function of the screened Poisson equation in $1$D with the screening length $\frac{1}{Q_{mn}}$. Solving the above differential equation gives:
\begin{align}
c_{mn}(x_\beta,x'_\beta) = 
\begin{cases}
\frac{1}{2 A}|x_\beta-x'_\beta| \,, & m = n = 0 \,, \\
-\frac{1}{2 A Q_{mn}} e^{-Q_{mn}|x_\beta-x'_\beta|} \,,& (m,n) \in \mathbb{Z}^2 \setminus (0,0) 
\end{cases} \,,
\end{align}
where the $(m,n) \neq (0,0)$ solution takes the form given for a doubly periodic charge distribution by Lennard-Jones and Dent \cite{lennard1928cohesion}. Now using the charge neutrality of the system and the symmetry relation $Q_{mn}=Q_{(-m)(-n)}$, together with the fact that the Green's function has the same periodicity as the system, we can obtain the expression for $\phi_0$ for the 2D periodic system as:
\begin{align}
\phi_0^{(2D)}(\bx) =& \frac{2\pi}{A} \bigg[\mathrm{sgn}(x_{\beta})\int_{\Omega} \mathrm{d} \bx' \, \left(\rho(\bx')+b(\bx',\bR)\right) x'_\beta + 2 \sum_{\substack{(m,n) \in \mathbb{Z}^2_+:\\Q_{mn} \leq Q_{mn}^\text{max}}} \Re \Bigg\{\frac{e^{i \bQ_{mn} \cdot \bx} \, e^{-Q_{mn}(x_\beta-\frac{L_\beta}{2}) \, \mathrm{sgn}(x_\beta)}}{Q_{mn}} \nonumber \\
&\times \int_{\Omega} \mathrm{d} {\bx'} \, e^{-i \bQ_{mn} \cdot \bx'} \left(\rho(\bx')+b(\bx',\bR)\right) e^{-Q_{mn} (\frac{L_\beta}{2}-x'_\beta) \, \mathrm{sgn}(x_\beta)} \Bigg\}\Bigg]\,, \quad \bx \in \partial \Omega_\beta, \beta \in \cI_o \,,
\end{align}
where $\mathrm{sgn}$ is the sign function, $\mathbb{Z}^2_+ = \{(m,n) \in \mathbb{Z}^2 : n > 0 \ \text{or} \ (n=0 \ \text{and} \ m>0)\}$, and $Q_{mn}^\text{max}$ is a truncation parameter. In the above expression, the first term is the dipole contribution, which causes the electrostatic potential to develop a step across the system, with the step size proportional to the dipole moment, as illustrated in Fig.~\ref{Fig:phi_illustration}. The remaining terms decay exponentially away from the system, so that the series converges given the compact support of the charge density.

It is worth reiterating that the expressions for $\phi_0$ derived above are valid on and beyond $\partial\Omega_\alpha$, $\alpha \in \cI_o$, where the charge density has vanished. For systems with vanishing electrostatic potential on the open boundaries, the proposed formulation reduces to that of previous works~\cite{Ghosh2017cluster,Ghosh2017extended}, which is recovered by setting $\bE$ and $\phi_0$ to zero. The developed formulation also accommodates fixed-potential boundary conditions, such as those imposed by external electrodes~\cite{ramakrishnan2025real}; on the open boundaries held at a specified potential, the Dirichlet value $\phi_0 + \bx \cdot \bE$ is replaced by that potential, and the surface integral vanishes, leaving the affine structure of $\mathcal{A}$ and the variational problem otherwise unchanged. Furthermore, the current formulation can be extended to semi-infinite systems such as surfaces by incorporating the bulk boundary condition approach \cite{bhowmik2026bulk}, wherein Dirichlet values matched to the electrostatic potential of the corresponding bulk crystal are prescribed at the bulk-facing boundary. Having developed this local variational formulation of the electrostatics, we employ it in the next section to construct the real-space pseudopotential Kohn--Sham DFT functional, from which we derive the electronic ground-state equations and the expressions for the ground-state energy, atomic forces, and stress tensor.
\section{Real-space DFT} \label{Sec:DFT}
Adopting the local real-space formulation of the electrostatic energy functional (Eq.~\eqref{Eqn:Electrostaticenergyclosedform}) derived in the previous section, we can now write the expression for the finite-temperature spin-restricted Kohn--Sham energy functional \cite{mermin1965thermal} on the unit cell $\Omega$ of the system as:
\begin{subequations} \label{Eqn:Energyfunctional}
\begin{align}
\mathcal{E}[\Psi,\bg;\bR,\bE] = & T_\text{s}[\Psi,\bg]+\mathcal{E}_\text{xc}[\rho, \nabla \rho]+\mathcal{E}_\text{el}[\rho;\bR,\bE]+\mathcal{E}_\text{nl}[\Psi,\bg;\bR]+\mathcal{E}_\text{ent}[\bg] \,,\\
 T_\text{s}[\Psi,\bg]= & -\sum_{\bk \in \Bz} w_\bk \sum_{n=1}^{N_b}  g_{n\bk} \int_\Omega \mathrm{d} \bx \, \psi^*_{n\bk}(\bx) \nabla^2 \psi_{n\bk}(\bx) \,,\\
 \mathcal{E}_\text{xc}[\rho,\nabla \rho] = & \int_\Omega \mathrm{d} \bx \, \varepsilon_{\text{xc}}(\rho(\bx),\nabla \rho(\bx)) \rho(\bx) \,,\\
 \mathcal{E}_\text{nl}[\Psi,\bg;\bR] = & 2\sum_{\bk \in \Bz} w_\bk \sum_{n=1}^{N_b} g_{n\bk} \sum_{I=1}^N \sum_{p=1}^{\mathcal{P}_I} \bigg|\int_\Omega \mathrm{d} \bx \, \psi^*_{n \bk}(\bx) \tilde{\chi}_{Ip\bk}(\bx,\bR_I) \bigg|^2 \,,\\
\mathcal{E}_\text{ent}[\bg]= & 2 k_B T \sum_{\bk \in \Bz} w_\bk \sum_{n=1}^{N_b} \left[g_{n\bk} \log g_{n\bk} + \left(1-g_{n\bk}\right) \log\left(1-g_{n\bk}\right) \right] \,,
\end{align}
\end{subequations}
where $T_\text{s}$ is the non-interacting kinetic energy, $\mathcal{E}_\text{xc}$ is the exchange-correlation energy within the generalized gradient approximation (GGA), $\mathcal{E}_\text{nl}$ is the nonlocal pseudopotential energy within the Kleinman--Bylander formalism \cite{kleinman1982efficacious}, and $\mathcal{E}_\text{ent}$ is the Fermi-Dirac electronic entropic energy. Here, $\Psi$ is the set of orbitals $\psi_{n\bk}$ that are Bloch-periodic in the directions $\alpha \in \cI_p$ and vanish on and beyond $\partial \Omega_\alpha, \alpha \in \cI_o$; $\bg$ is the set of occupation numbers $g_{n\bk} \in [0,1]$; $\Bz$ is the discrete set of electronic wavevectors $\bk$ sampling the Brillouin zone, with weights $w_\bk$ satisfying $\sum_{\bk \in \Bz} w_\bk = 1$; $N_b$ is the number of Kohn--Sham bands; $\psi_{n\bk}^*$ is the complex conjugate of $\psi_{n\bk}$; $\varepsilon_{\text{xc}}$ is the exchange-correlation energy per electron;
\begin{align}
\rho(\bx) = 2\sum_{\bk \in \Bz} w_\bk \sum_{n=1}^{N_b} g_{n\bk} |\psi_{n\bk}(\bx)|^2\label{Eqn:electrondensity}
\end{align}
is the electron density; $\mathcal{P}_I$ is the number of projectors for atom $I$, with $\tilde{\chi}_{Ip\bk}$ the corresponding Bloch-periodically mapped normalized projectors; $k_B$ is the Boltzmann constant; and T is the electronic temperature.

The electronic ground state of the system is the solution of the following minimization problem:
\begin{subequations}
\begin{align}
\min_{\Psi,\bg} \mathcal{E}[\Psi,\bg;\bR,\bE] \,,
\end{align}
subject to the constraints of orthonormality of the orbitals and conservation of the total number of electrons $N_e$ in $\Omega$, i.e.,
\begin{align}
\int_\Omega \mathrm{d} \bx \, \psi^*_{n\bk}(\bx) \psi_{m\bk}(\bx) = \delta_{nm} \,, \quad \forall \, n,m \in \{1,2,\ldots, N_b\},\ \bk \in \Bz\,, \quad 2 \sum_{\bk \in \Bz} w_\bk \sum_{n=1}^{N_b} g_{n\bk} = N_e \,.
\end{align}
\end{subequations}
 The corresponding Euler--Lagrange equations are given by:
 \begin{subequations}
\begin{align}
\left(-\frac{1}{2} \nabla^2 + V_{\text{xc}} + \phi + V_{\text{nl}\bk} \right) \psi_{n\bk}(\bx) &= \lambda_{n\bk} \psi_{n\bk}(\bx)\,, \quad \bx \in \Omega \label{Eqn:eigenvalue}\\
g_{n\bk} &= \left(1+e^{\frac{\lambda_{n\bk}-\mu}{k_B T}} \right)^{-1} \,,\quad \mu \, \text{determined by} \,\, 2 \sum_{\bk \in \Bz} w_\bk \sum_{n=1}^{N_b} g_{n\bk} = N_e \label{Eqn:occupation}\,,
\end{align}
\end{subequations}
where $V_{\text{xc}}$ is the exchange-correlation operator, $\phi$ is the solution of the  Poisson equation (Eq.~\eqref{Eqn:Poisson}), $V_{\text{nl}\bk}$ is the nonlocal pseudopotential operator, $\lambda_{n\bk}$ is the Kohn--Sham eigenvalue, and $\mu$ is the chemical potential. Notably, the Kohn--Sham Hamiltonian in Eq.~\eqref{Eqn:eigenvalue} retains the same form as in the absence of the applied electric field, since the field enters entirely through the electrostatic potential $\phi$; this is in contrast to conventional treatments \cite{kunc1983external,meyer2001ab}, where the Hamiltonian is explicitly modified by an additional field-dependent term. Because the Kohn–Sham operator in Eq.~\eqref{Eqn:eigenvalue} depends on its own eigenfunctions (through $\rho$), the equation is solved self-consistently. In each iteration, the electron density (Eq.~\eqref{Eqn:electrondensity}) is updated using the eigenfunctions and occupation numbers (Eq.~\eqref{Eqn:occupation}) and is in turn used to update the exchange-correlation and electrostatic (Eq.~\eqref{Eqn:Poisson}) potentials, thereby forming the operator for the next iteration. The self-consistent orbitals and occupation numbers constitute the electronic ground state of the system.

Substituting the minimizers $\Psi$ and $\bg$ and maximizer $\phi$ in Eq.~\eqref{Eqn:Energyfunctional} gives the following Harris-Foulkes type \cite{harris1985simplified,foulkes1989tight} expression for the ground-state energy:
\begin{align}
\mathcal{E}(\bR,\bE) =& 2 \sum_{\bk \in \Bz} w_\bk \sum_{n=1}^{N_b}  g_{n\bk} \lambda_{n\bk} + \mathcal{E}_\text{xc} -\int_\Omega \mathrm{d} \bx \, V_\text{xc}(\rho(\bx),\nabla \rho(\bx)) \rho(\bx) + \frac{1}{2} \int_\Omega \mathrm{d} \bx \, \Big((b(\bx,\bR)-\rho(\bx)) \phi(\bx,\bR) + \bx \cdot \bE \nonumber \\
&  \left(\rho(\bx)+b(\bx,\bR)\right) \Big)  + \frac{1}{2} \int_\Omega \mathrm{d} \bx \, \Big( V_c(\bx,\bR) \left(b(\bx,\bR)+\tilde{b}(\bx,\bR)\right) - \sum_{I=1}^M \tilde{b}_I(\bx,\bR_I) \tilde{V}_I(\bx,\bR_I) \Big) +\mathcal{E}_\text{ent} \,,
\end{align}
where $\tilde{b} = \sum_{I=1}^M \tilde{b}_I$ is the total reference ionic pseudocharge density, with $\tilde{b}_I = -\frac{1}{4 \pi} \nabla^2 \tilde{V}_I$ the spherically symmetric and localized reference ionic pseudocharge density corresponding to the reference ionic pseudopotential $\tilde{V}_I$ of the $I^\text{th}$ ion; $V_c = \sum_{I=1}^M(\tilde{V}_I-V_I)$ is the correction potential. The fifth term in the above expression is the self-interaction and overlap-correction energy $\mathcal{E}_\text{sc}$. It is important to note that the above expression differs from that of previous work~\cite{Ghosh2017extended} only by the electric-field-dependent contribution, and therefore reduces to it in the absence of the field.

Differentiating the ground-state energy of $\Omega$ with respect to the atomic position $\bR_{I}$ of atom $I$ and invoking the Hellmann--Feynman theorem \cite{feynman1939forces}, the atomic force on atom $I$ is obtained as:
\begin{align}
\mathbf{f}_I(\bR,\bE) =& -\frac{\partial \mathcal{E}(\bR,\bE)}{\partial \bR_{I}} \nonumber \\
=& -\sum_{I'} \int_\Omega \mathrm{d}\bx \, b_{I}(\bx,\bR_{I'}) \nabla \phi(\bx,\bR) -\sum_{I'} \frac{1}{2} \int_\Omega \mathrm{d} \bx \, \nabla V_c(\bx,\bR)\left(b_{I}(\bx,\bR_{I'})+\tilde{b}_{I}(\bx,\bR_{I'}) \right) \nonumber \\
& + \sum_{I'}\frac{1}{2} \int_\Omega \mathrm{d} \bx \, \nabla {V_c}_{I}(\bx,\bR_{I'}) \left(b(\bx,\bR)+\tilde{b}(\bx,\bR)\right) -4 \sum_{\bk \in \Bz} w_\bk \sum_{n=1}^{N_b} g_{n\bk} \sum_{p=1}^{\mathcal{P}_I} \Re \bigg[\int_\Omega \mathrm{d} \bx \, \psi^*_{n\bk}(\bx) \tilde{\chi}_{Ip\bk}(\bx,\bR_{I}) \nonumber \\
&\times \int_\Omega \mathrm{d} \bx \, \tilde{\chi}^*_{Ip \bk}(\bx,\bR_{I}) \nabla \psi_{n\bk}(\bx) \bigg] \,,
\end{align}
where $I'$ denotes atom $I$ together with all its periodic images, $V_{cI} = \tilde{V}_I - V_I$, and $\Re [\cdot]$ denotes the real part. In the expression, the applied electric field enters only implicitly through $\phi$; there is no explicit field contribution, so the expression retains the same form in the absence of electric field too. The contribution of the self-interaction term is eliminated using the spherical symmetry of the pseudocharge density. The expression thus coincides with that of Ghosh~et~al.~\cite{Ghosh2017extended}, apart from the simplification of the self-interaction terms.

The stress tensor follows from the strain derivative of the ground-state energy; extending the zero-field result of Sharma~et~al.~\cite{sharma2018calculation} to include the applied field gives:
\begin{align}
\sigma_{\alpha \beta}(\bR,\bE) =\sigma_{\alpha \beta}(\bR,\mathbf{0}) + \delta_{\alpha \beta} \frac{1}{2|\Omega|} \int_\Omega \mathrm{d} \bx \, \left(\rho(\bx)+b(\bx,\bR)\right) \bx \cdot  \bE \,,
\end{align}
where $\sigma_{\alpha \beta}(\bR,\mathbf{0})$ is the zero-field stress tensor whose expression is given in Eq.~(22) of Ref.~\cite{sharma2018calculation}. Since the applied electric field is uniform and acts only along the open directions, its contribution to the stress is isotropic, appearing only in the diagonal components.
\section{Finite-difference implementation} \label{Sec:implementation}
We now discuss the implementation of the above derived electrostatics---in particular, the Dirichlet boundary conditions---for isolated and partially periodic systems in the large-scale parallel GPU-accelerated \cite{sharma2023gpu,jing2025gpu} SPARC electronic structure code \cite{xu2021sparc,zhang2024sparc}. Within the code, all quantities are represented on a 3D finite-difference grid which is obtained by uniformly discretizing the unit cell along $\bL_1$, $\bL_2$, and $\bL_3$. Derivatives are evaluated using a high-order centered finite-difference scheme, and integrations are performed using the trapezoidal rule. The pseudocharge density contribution from each atom is obtained by applying the discrete Laplacian to the atom's pseudopotential in a matrix-free manner \cite{suryanarayana2014augmented}, evaluated only at the grid points lying within the pseudocharge cutoff region around the atom. The electron density is computed self-consistently, starting from a superposition of atomic electron densities and updated at each iteration using the eigenpairs of the Hamiltonian obtained via the Chebyshev-filtered subspace iteration (CheFSI) technique \cite{zhou2006self,zhou2006parallel}. The convergence of the self-consistent-field (SCF) cycle is accelerated using the restarted variant \cite{pratapa2015restarted} of the Periodic Pulay mixing scheme \cite{Banerjee2016PeriodicPulay} with a real-space preconditioner \cite{realspaceprecond}. In each SCF step, Poisson equation is solved using the Alternating Anderson--Richardson (AAR) iterative method \cite{suryanarayana2019alternating, pratapa2016anderson}, with the electrostatic potential from the previous step used as the initial guess. On distributed-memory computing architectures, the equation is solved in parallel by distributing the discretized domain equally among the processors and employing the message passing interface (MPI) standard to perform collective operations and point-to-point communications.
 
To impose the derived Dirichlet boundary conditions on the electrostatic potential, we first identify the grid points within each processor domain whose discrete-Laplacian matrix-vector product requires a contribution from ghost nodes—the fictitious grid points lying outside the domain boundary along the open directions. We then evaluate, at each such ghost node, the potential due to the total charge density and any applied electric field, using the analytical expressions derived in Sec.~\ref{Sec:Electrostatics}. For every identified grid point in the domain, the net ghost-node contribution is obtained by weighting these potential values with the finite-difference coefficients of the discrete second-order derivatives along the corresponding open directions and summing the results. This contribution is subtracted from the right-hand side of the Poisson equation, and the resulting system is solved using the same machinery described above. 

The number of ghost nodes at which these analytical expressions must be evaluated is $N_o \sum_{\alpha \in \cI_o} N_{\alpha}$, where $N_o$ is the finite-difference order and $N_\alpha$ is the number of grid points on each $\partial \Omega_\alpha$. Using this, we can estimate the scaling of the ghost-node contribution to the electrostatic potential with the total number of grid points $N_d$ for each geometry. For the isolated system, the separability of the $\bx$- and $\bx'$-dependent terms in $\phi_0^{(0D)}$ allows the expression to be evaluated with $\mathcal{O}(N_d)$ scaling. For the 1D periodic system, the same $\mathcal{O}(N_d)$ scaling holds because the first series of $\phi_0^{(1D)}$ is separable in $\bx$ and $\bx'$, while the $K_0$ term in the second series, though dependent on both $\bx$ and $\bx'$, need only be evaluated in the plane, keeping the combined cost linear in $N_d$. For the 2D periodic system, $\mathcal{O}(N_d)$ scaling likewise holds, since the $\bx$- and $\bx'$-dependent terms in $\phi_0^{(2D)}$ separate for both the $Q_{mn}=0$ and $Q_{mn} > 0$ terms. These evaluations parallelize naturally within the domain-decomposition framework for the Poisson solution in SPARC: the source-dependent integrals---the spherical multipole moments for the isolated system, the cylindrical multipole moments together with the axial Fourier coefficients for the 1D periodic system, and the in-plane Fourier coefficients for the 2D periodic system---are first accumulated as local partial sums over each processor's grid points at a cost of $\mathcal{O}(N_d/N_p)$, where $N_p$ is the number of processors, and then combined through collective reductions over the appropriate Cartesian communicators. Each processor subsequently evaluates the boundary potential independently at the ghost nodes it owns, combining the precomputed moments and coefficients with the local, position-dependent factors, with no further communication.

The evaluation of these boundary potentials further benefits from geometry-specific numerical choices. For the isolated system, the boundary conditions are computed using real spherical harmonics, which avoids complex arithmetic in their evaluation and yields a significant speedup. For the 1D periodic system, the required zeroth-order modified Bessel function of the second kind is efficiently evaluated using the routine from Numerical Recipes \cite{press2007numerical}. In all cases, since the boundary potentials depend only on the total charge density and the applied field, they are computed once per SCF iteration and reused across the AAR iterations. Beyond the boundary conditions, the additional electric-field-dependent terms in the energy and stress expressions derived in the previous section are also implemented in SPARC.
\section{Numerical validation} \label{Sec:numericaltests}
\begin{figure}[!htbp]
    \centering
    \includegraphics[width=\textwidth,keepaspectratio=true]{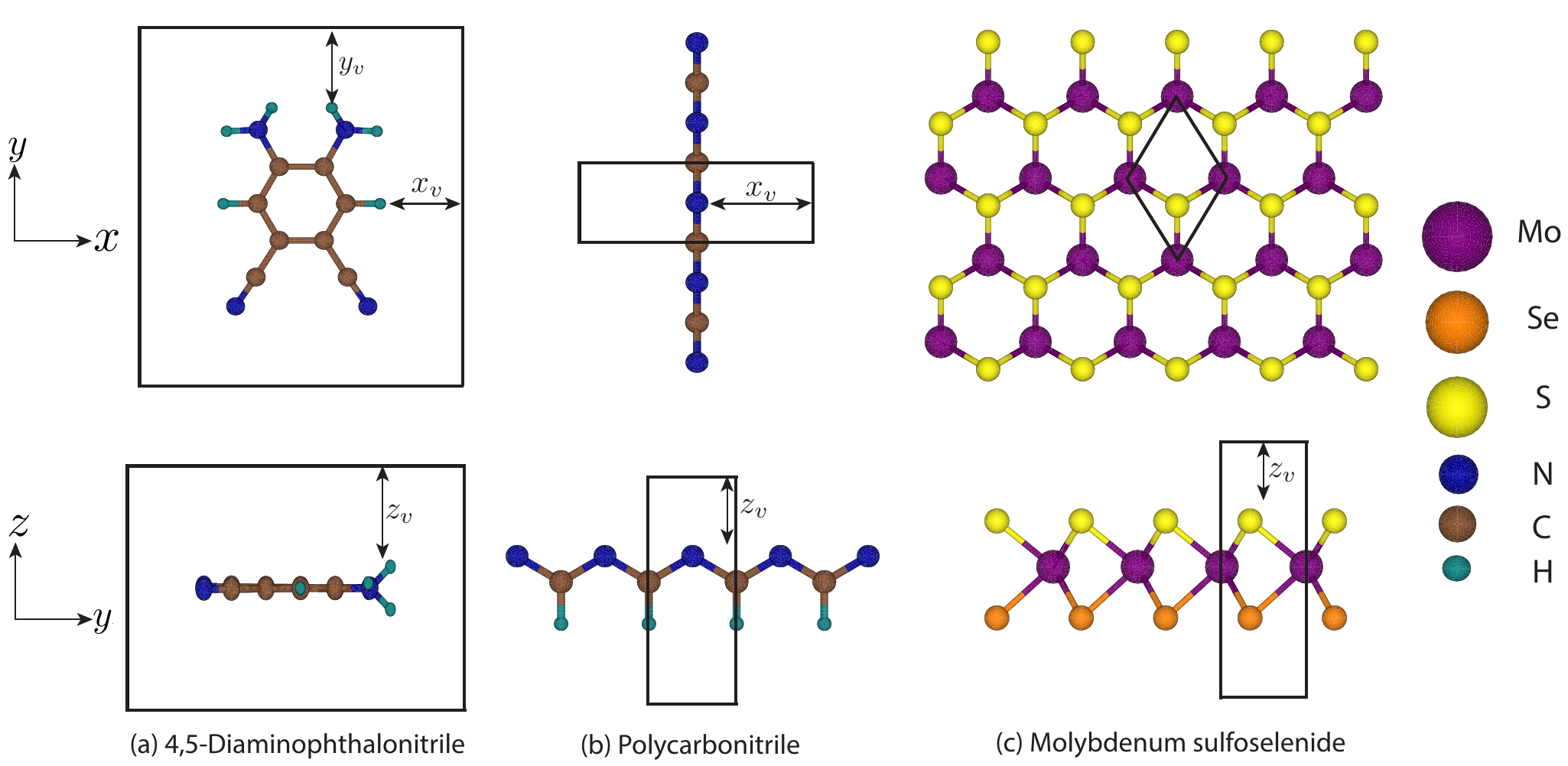}
    \caption {Illustration showing the atomic arrangement and the unit cell with vacuum for (a) the 4,5-diaminophthalonitrile molecule, (b) the polycarbonitrile wire, and (c) the molybdenum sulfoselenide Janus monolayer. The top and bottom rows show projections onto the $x$--$y$ and $y$--$z$ planes, respectively. Here, $x_v$, $y_v$, and $z_v$ denote the vacuum size in the $x$, $y$, and $z$ directions, respectively. Atomic species are indicated in the legend on the right.}
    \label{Fig:lowDexamples}
\end{figure}

We now test the efficiency and accuracy of the developed real-space formulation and implementation. For this purpose, we select a $4,5$-diaminophthalonitrile (C$_8$H$_6$N$_4$) molecule, a polycarbonitrile $(\text{CHN})_x$ wire, and a molybdenum sulfoselenide (MoSSe) Janus monolayer, as representative examples of isolated, 1D periodic, and 2D periodic systems, respectively; each was specifically chosen for exhibiting appreciable multipole moments along at least one of its open directions. Fig.~\ref{Fig:lowDexamples} illustrates the atomic arrangement and the unit cell with vacuum for these systems: C$_8$H$_6$N$_4$ consists of a central benzene ring substituted with two amino groups and two nitrile groups; (CHN)$_x$ is a planar zigzag polymer constructed using the periodic repetition of a 3-atom unit cell along the $y$-direction; and MoSSe has a honeycomb structure constructed using the periodic repetition of a 3-atom unit cell in the $x$--$y$ plane. For the molecule and the wire, the vacuum size is defined as the maximum of the vacuum along any of the open directions.

In all the simulations, we employ a twelfth-order-accurate finite-difference discretization of the differential operators, Monkhorst-Pack grid \cite{monkhorst1976special} for the discretization of the Brillouin zone, optimized norm-conserving Vanderbilt pseudopotentials (ONCV) \cite{hamann2013optimized} from the Shojaei-Pask-Medford-Suryanarayana (SPMS) \cite{spms} set, the Perdew-Burke-Ernzerhof (PBE) \cite{perdew1996generalized} variant of the GGA semilocal exchange-correlation functional, and Fermi-Dirac smearing of $1$ mHa. Furthermore, in all simulations of the isolated and 1D periodic systems, we use equal unit-cell lengths along the open directions, so as to respect the symmetry of the spherical and cylindrical multipole moments present in their respective boundary conditions. Unless specified otherwise, the mesh sizes and $k$-point grids adopted, as determined from convergence tests are: $0.11$ Bohr and no $k$-point for the molecule, $0.13$ Bohr and $6$  for the wire, and $0.13$ Bohr and $8\times8$ for the monolayer. Lastly, the errors reported for the atomic forces and stresses correspond to the maximum absolute error across any of their respective components.
\subsection{Convergence}
\begin{figure}[!htbp]
    \centering
    \subfloat[4,5-diaminophthalonitrile molecule\label{Fig:molecule_vacuum}]{
        \includegraphics[width=0.49\textwidth,keepaspectratio=true]{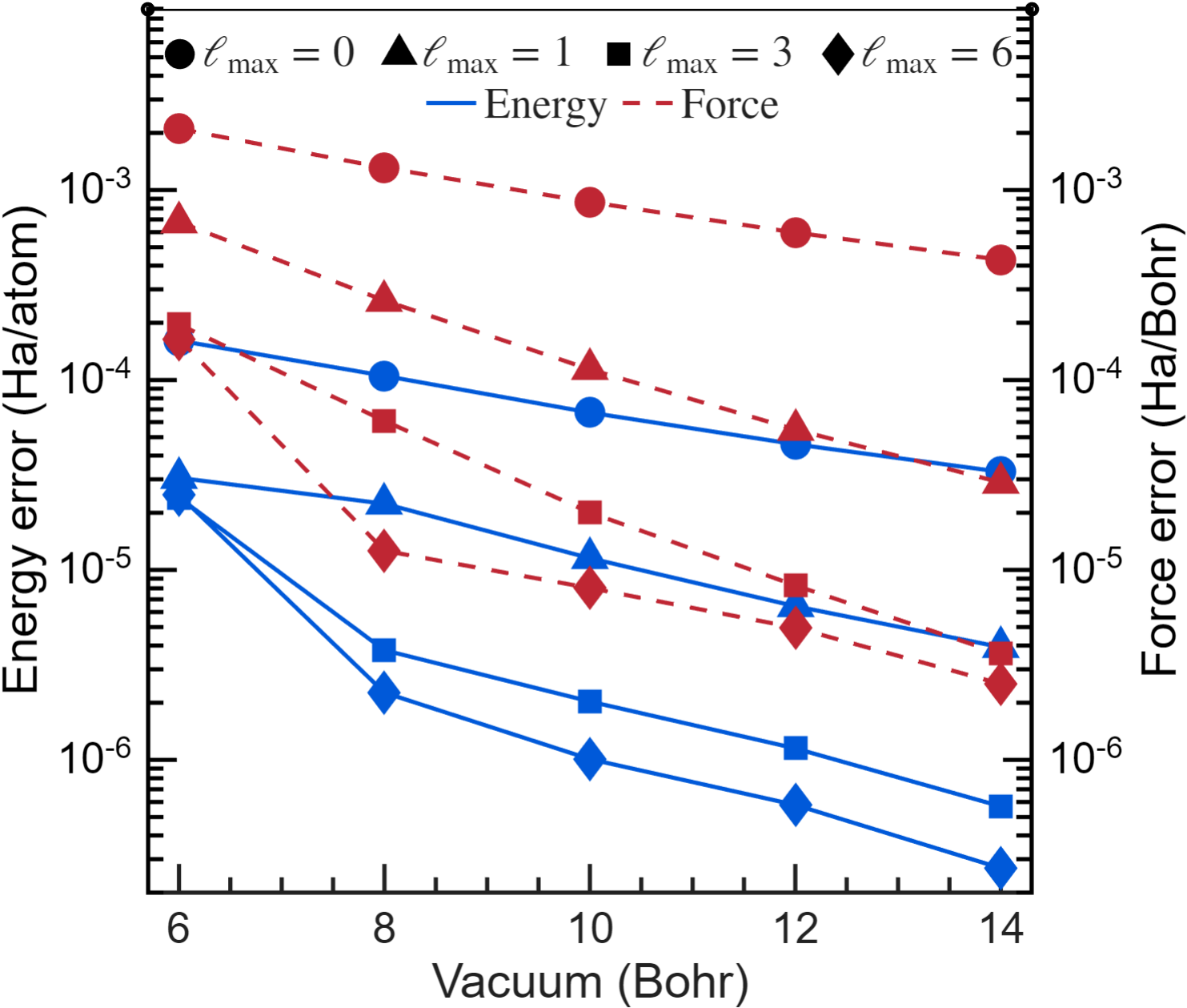}
    }
    \subfloat[Polycarbonitrile wire \label{Fig:wire_vacuum}]{
        \includegraphics[width=0.49\textwidth,keepaspectratio=true]{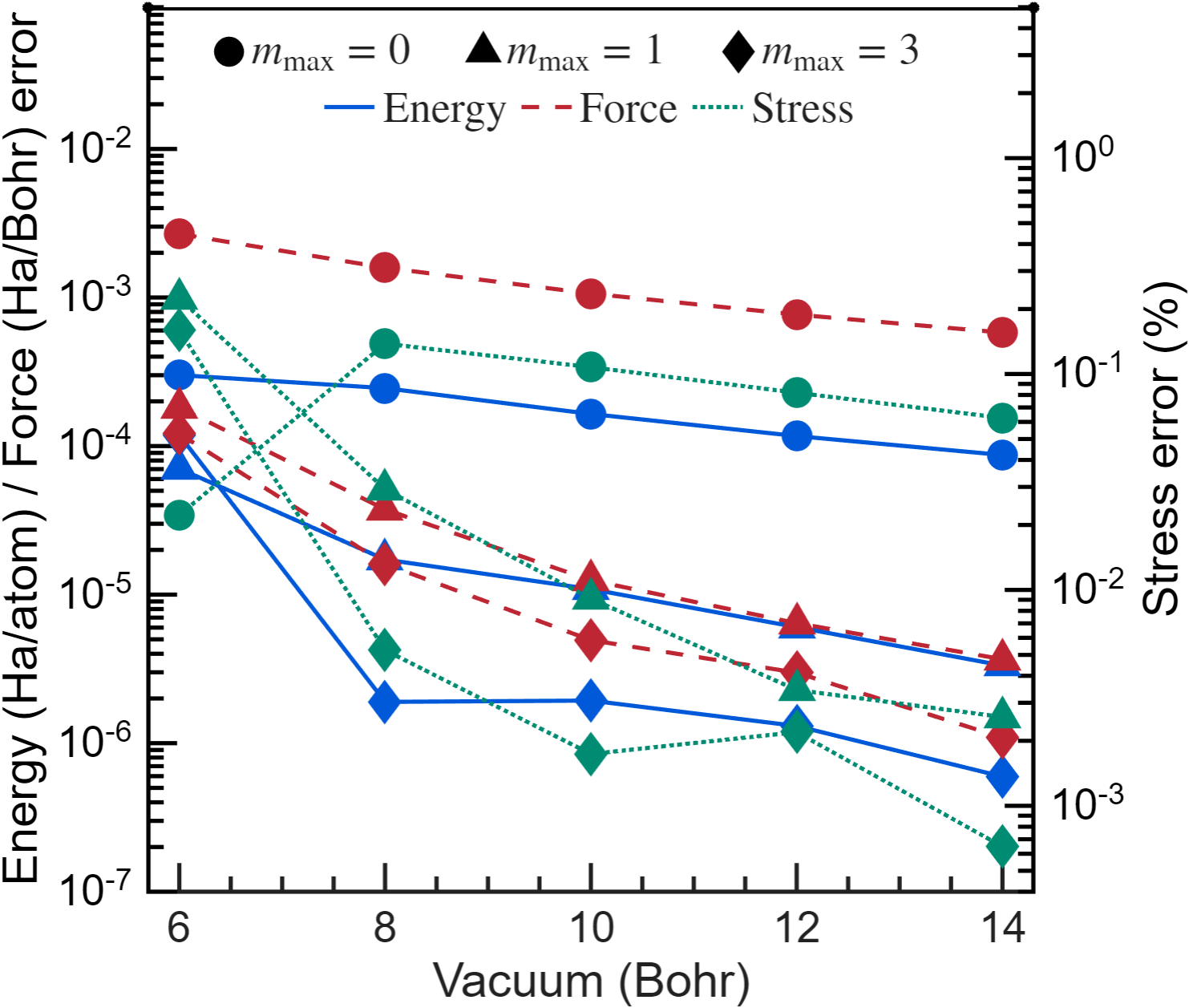}
    }\\
    \subfloat[Molybdenum sulfoselenide monolayer\label{Fig:surface_vacuum}]{
        \includegraphics[width=0.49\textwidth,keepaspectratio=true]{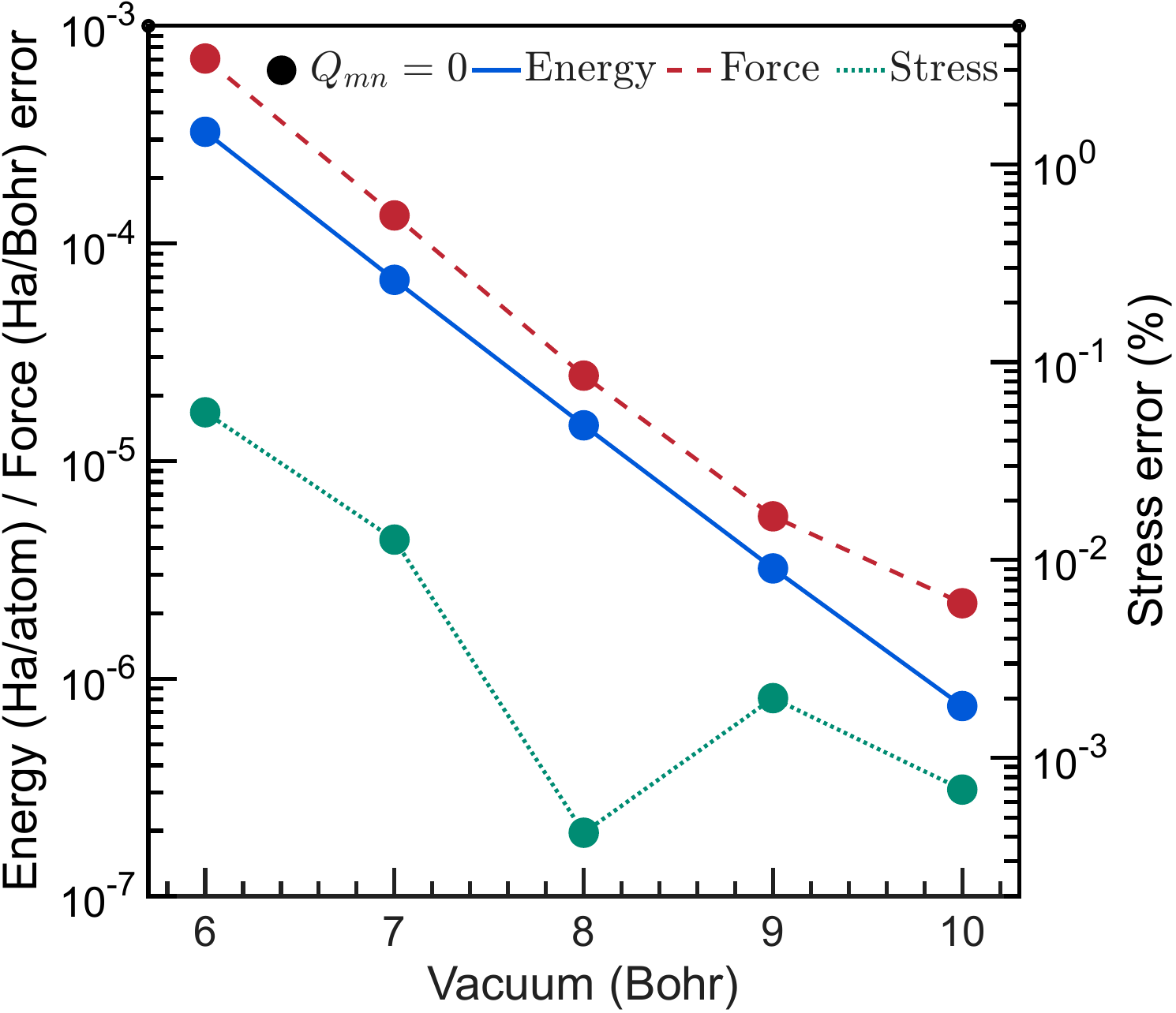}
    }
    \caption{Convergence of the energy (solid blue), forces (dashed red), and stresses (dotted green) with respect to vacuum size for (a) the 4,5-diaminophthalonitrile molecule, (b) the polycarbonitrile wire, and (c) the molybdenum sulfoselenide monolayer, at different truncation parameters. The contributions of $n > 0$ and $Q_{mn} > 0$ is negligible for the wire and monolayer, respectively, and therefore omitted. } \label{Fig:vacuumConv}
\end{figure}

We now examine the convergence of energy, atomic forces, and stresses with respect to vacuum size and truncation parameters for the systems described above. For this study, a mesh spacing of $0.2$ Bohr is employed and an electric field of $2 \times 10^{-3}$ Ha/(e$\cdot$Bohr) is applied along each open direction of the system. All reference calculations employ a $16$ Bohr vacuum size, with truncation parameters of $\ell_\text{max}=9$ for the molecule, $(m_\text{max},n_\text{max}) = (4,2)$ for the wire, and $Q_{mn}^\text{max} = 0.11$ Bohr$^{-1}$ for the monolayer; these are verified to yield results that are converged well below the smallest reported errors. Fig.~\ref{Fig:vacuumConv} shows the convergence of energy and forces for all the systems as well as the convergence of stresses for the wire and monolayer.

For the molecule, both energy and forces converge exponentially and monotonically with vacuum size at each $\ell_\text{max}$, with the convergence rate increasing for larger $\ell_\text{max}$, as shown in Fig.~\ref{Fig:molecule_vacuum}. Convergence with respect to the truncation parameter $\ell_\text{max}$ is also rapid; each increment reduces the error by nearly an order of magnitude, though returns diminish beyond $\ell_\text{max}=3$. At $\ell_\text{max} = 6$ and $10$ Bohr vacuum, the errors in energy and forces reach $1\times10^{-6}$ Ha/atom and $8\times10^{-6}$ Ha/Bohr, respectively. For the wire, each of the energy, forces, and axial stress converges exponentially with vacuum size at each $m_\text{max}$ (with $n_\text{max}=0$), with the convergence rate increasing as $m_\text{max}$ becomes larger, as shown in Fig.~\ref{Fig:wire_vacuum}. Each increment of $m_\text{max}$ lowers the error substantially, reaching $2\times10^{-6}$ Ha/atom in energy, $5\times10^{-6}$ Ha/Bohr in force, and $2 \times 10^{-3} \%$ in axial stress at $m_\text{max} = 3$ and $10$ Bohr vacuum. The $n > 0$ (Bessel) contributions are negligible here, as the short axial period of the wire renders $K_0$ exponentially small. Finally, for the monolayer as well, energy, forces, and stresses converge exponentially with vacuum size, reaching an error of $7\times10^{-7}$ Ha/atom, $2\times10^{-6}$ Ha/Bohr, and $7\times 10^{-4}\%$ by $10$ Bohr vacuum, respectively, as evident from Fig.~\ref{Fig:surface_vacuum}. The $Q_{mn} \neq 0$ terms are exponentially suppressed likely due to the small in-plane lattice lengths of the monolayer, so that the leading dipole term dominates the boundary potential. Across all three systems, the forces converge at rates comparable to the energy, albeit with somewhat larger errors at matched truncation parameters and vacuum size, likely due to their dependence on the gradients of the electrostatic quantities.

While negligible for the systems above, the nonzero Fourier components ($n > 0$ for the wire and $Q_{mn} > 0$ for the monolayer) can play an important role in accelerating convergence with vacuum size for certain partially periodic systems, as demonstrated in Appendix~\ref{Sec:AppendixA}. Overall, the developed framework demonstrates an exponential convergence of the energy, forces, and stresses with vacuum size, reaching the $\sim \mu$Ha level in energy and forces (and $\sim \! 10^{-3} \%$ in stresses) at a vacuum of 10 Bohr across all three geometries. For the remainder of the simulations, we employ a vacuum size of $10$ Bohr for all the systems considered and truncation parameters of $\ell_\text{max}=6$ for the molecule, $(m_\text{max},n_\text{max})=(3,0)$ for the wire, and $Q_{mn}^\text{max}=0$ for the monolayer.
\subsection{Accuracy}
\begin{figure}[!htbp]
    \centering
    \subfloat[Computed energy and its quadratic fit\label{Fig:energyfit}]{
        \includegraphics[width=0.49\textwidth,keepaspectratio=true]{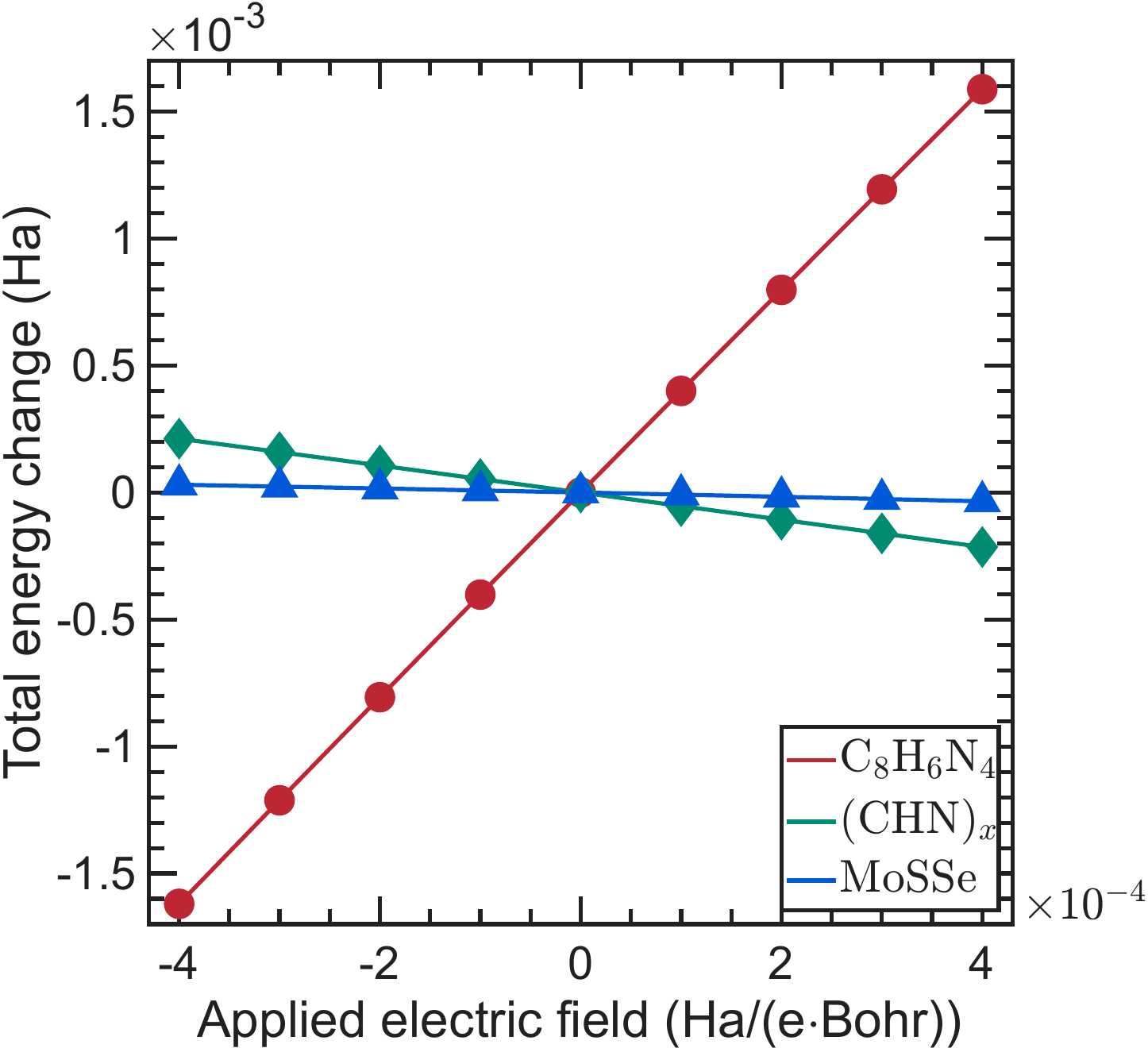}}
    \hfill
    \subfloat[Computed dipole and energy derivative \label{Fig:dipoleconsistency}]{
        \includegraphics[width=0.49\textwidth,keepaspectratio=true]{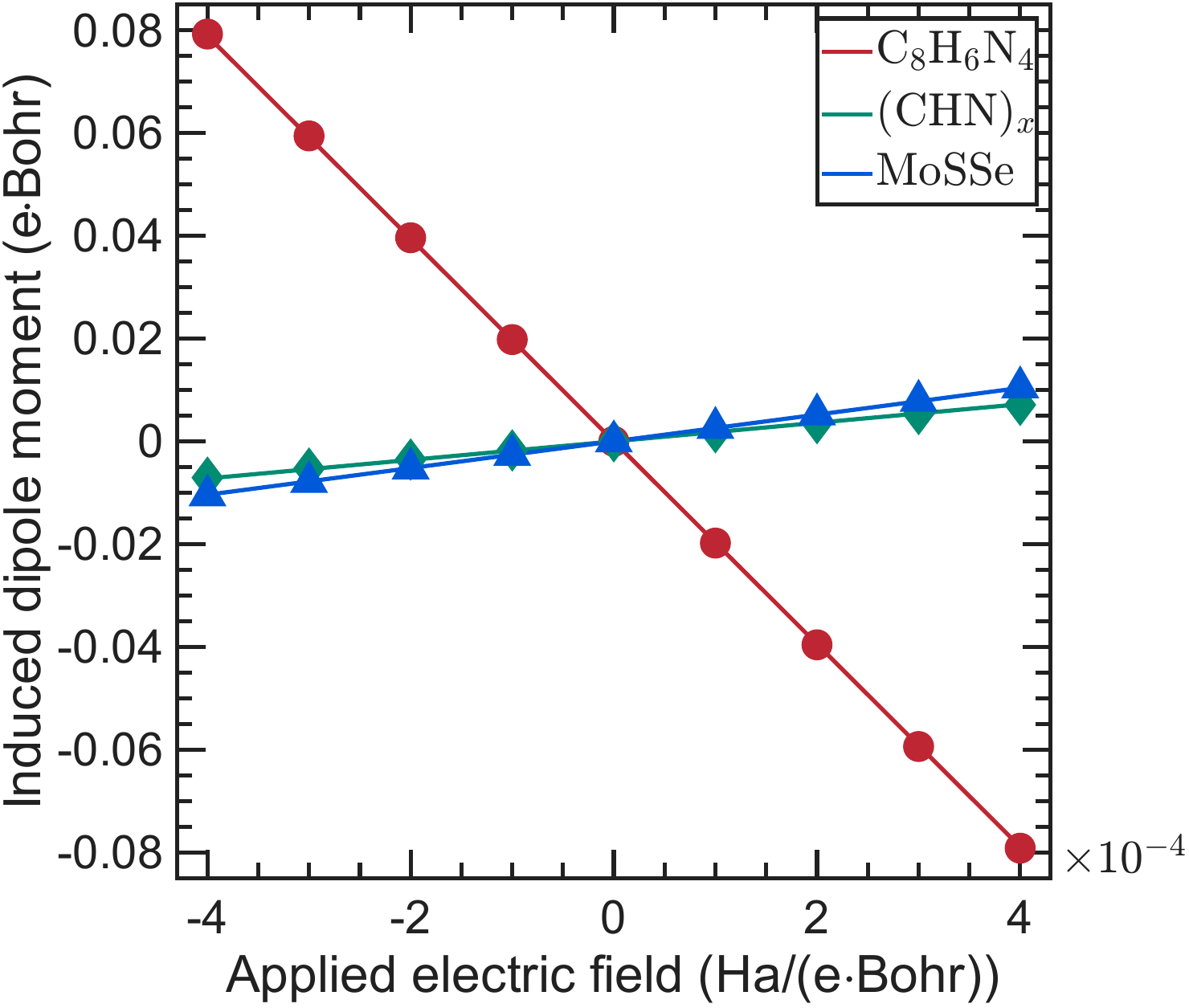}}
    \caption{Consistency of the energy and dipole moment for the 4,5-diaminophthalonitrile (C$_8$H$_6$N$_4$) molecule, polycarbonitrile ($(\text{CHN})_x$) wire, and molybdenum sulfoselenide (MoSSe) monolayer. (a) Change in computed energy (markers) with electric field and its quadratic fit (curves). (b) Change in the computed dipole moment (markers) with the applied field, compared with the derivative of the fitted energy with respect to the field (curves).}
    \label{Fig:consistency}
\end{figure}

We now verify the accuracy of the framework by first performing an internal consistency test based on the energy--dipole relation, followed by a comparison against the established plane-wave DFT code Quantum ESPRESSO \cite{giannozzi2009quantum}. To check the consistency of the dipole moment 
\begin{align}
\boldsymbol{\mu} = -\frac{\partial \mathcal{E}}{\partial {\bE}}
\end{align}
with the energy, we compare the dipole moment computed from the first moment of the total charge density against the negative numerical derivative of the total energy with respect to the applied electric field. For each system, the ground-state energy and dipole moment are computed over a range of electric fields applied along the $z$-direction. Fig.~\ref{Fig:energyfit} shows the change in computed energy (markers) with electric field together with its quadratic fit (curves), and Fig.~\ref{Fig:dipoleconsistency} shows the change in computed dipole moment (markers) with electric field together with the derivative of the fitted energy with respect to the electric field (curves). The maximum difference obtained between the two dipole moments are: $3.49\times10^{-4}$ e$\cdot$Bohr for the molecule, $1.54\times10^{-4}$ e$\cdot$Bohr for the wire, and $1.31\times10^{-4}$ e$\cdot$Bohr for the monolayer. Since the two dipole evaluations agree closely across all field strengths for all three systems, it establishes the consistency of the dipole moment with the energy within the developed framework.

We next compare energy, atomic forces, and polarization density along the $z$-direction
\begin{align}
P_z=\frac{\mu_z}{|\Omega|}
\end{align}
computed by SPARC against those from Quantum ESPRESSO for the three systems, both with and without an applied electric field. To ensure a rigorous comparison, the plane-wave calculations were carefully converged. The molecule required a $40$ Bohr vacuum with wavefunction and density truncations of $193.5$ Ha and $774$ Ha, respectively; the wire and monolayer required $80$ Bohr and $20$ Bohr of vacuum, respectively, both with wavefunction and  density truncations of $138.5$ Ha and $554$ Ha, respectively. As evident from Table~\ref{Tab:sparc_qe}, the energy agrees to within $\sim\!10^{-5}$--$10^{-4}$ Ha/atom, the forces to within $\sim\!10^{-5}$ Ha/Bohr, and the polarization density to $\sim \!0.1 \%$ across all systems, confirming the accuracy of the framework. The energy error is largest for the monolayer, which we attribute to the nonlinear core correction present in the SPMS pseudopotentials of its constituent elements. To verify this, we replace the SPMS set with the SG15 pseudopotentials~\cite{hamann2013optimized}, which omit the core correction, and observe that the energy error for the monolayer reduces to $4.16\times10^{-6}$~Ha/atom, confirming the accuracy of the framework for the 2D periodic systems as well. Since the stress for these systems is not available from the existing DFT codes, we instead validate it against finite-difference derivatives of the energy. For the wire, the error in the axial stress is $0.097 \%$ at $\bE = \mathbf{0}$ and $0.098 \%$ at $\bE = 0.001\,\hat{\mathbf{e}}_z$ Ha/(e$\cdot$Bohr). For the monolayer, the maximum absolute error among the in-plane stresses is $0.075\%$ at $\bE = \mathbf{0}$ and $0.076 \%$ at $\bE = 0.01\,\hat{\mathbf{e}}_z$ Ha/(e$\cdot$Bohr). These small errors validate the accuracy of the stresses
within the developed framework. It is important to note that the errors are essentially unchanged between the zero-field and finite-field calculations, demonstrating that the introduction of the electric field does not degrade the accuracy of the framework.

Finally, we emphasize that the additional cost of evaluating the Dirichlet boundary conditions is negligible---less than $0.01\%$ of the total Poisson-solve time---so that the performance of the electrostatic solver is effectively unchanged from that of the previous implementation in SPARC. Consequently, the developed framework retains the parallel scalability of SPARC while extending it to accurately and efficiently study systems with non-vanishing electrostatic potential on the open boundaries.
\begin{table}[!htbp]
\centering
\caption{Comparison of SPARC and Quantum ESPRESSO for the 4,5-diaminophthalonitrile molecule, polycarbonitrile wire, and molybdenum sulfoselenide monolayer, with and without an applied electric field. Reported are the maximum absolute differences in energy, atomic forces, and polarization density; values in parentheses are the actual values from SPARC.}
\label{Tab:sparc_qe}
\begin{tabular}{llccc}
\toprule
System & Electric field & \shortstack{Energy difference\\(Ha/atom)}
& \shortstack{Force difference\\(Ha/Bohr)}
& \shortstack{Polarization density difference\\(e$\cdot$Bohr$^{1-d}$)} \\
\midrule
\multirow{2}{*}{C$_8$H$_6$N$_4$}
& $\mathbf{0}$           & $1.18\times10^{-5}$ & $5.10\times10^{-5}$ & $8.07\times10^{-3}$ ($4.01$) \\
& $0.001\,\hat{\mathbf{e}}_z$ & $1.23\times10^{-5}$ & $4.84\times10^{-5}$ & $7.83\times10^{-3}$ ($3.81$) \\
\midrule
\multirow{2}{*}{(CHN)$_x$}
& $\mathbf{0}$           & $1.24\times10^{-5}$ & $4.43\times10^{-5}$ & $9.60\times10^{-5}$ ($0.12$) \\
& $0.001\,\hat{\mathbf{e}}_z$ & $6.82\times10^{-6}$ & $1.66\times10^{-5}$ & $7.90\times10^{-5}$ ($0.12$) \\
\midrule
\multirow{2}{*}{MoSSe}
& $\mathbf{0}$           & $1.06\times10^{-4}$ & $1.39\times10^{-5}$ & $4.04\times10^{-6}$ ($0.0025$) \\
& $0.01\,\hat{\mathbf{e}}_z$  & $1.06\times10^{-4}$ & $9.21\times10^{-6}$ & $1.51\times10^{-5}$ ($0.0105$) \\
\bottomrule
\end{tabular}
\end{table}
\section{Applications}\label{Sec:applications}
We now apply our developed and numerically tested framework to compute the static polarizabilities of the above described molecule, wire, and monolayer, as well as the piezoelectric coefficients of the wire and monolayer. For these calculations, we use the relaxed geometries obtained from force and stress relaxation with tolerances of $5\times10^{-4}$ Ha/Bohr and $0.1$ GPa, respectively. The atomic positions of the relaxed molecule are provided in the accompanying data. The equilibrium geometry of the wire has lattice parameter $L_y = 4.320$ Bohr, C--N and C--H bond lengths of $2.513$ Bohr and $2.108$ Bohr, respectively, and C--N--C and N--C--H bond angles of $118.52^\circ$ and $120.63^\circ$, respectively. The equilibrium geometry of the monolayer has a lattice parameter $L_x=L_y=6.140$ Bohr and the out-of-plane displacements of $2.889$ Bohr and $3.224$ Bohr for S and Se atoms, respectively. In all the simulations, we employ a vacuum size of $10$ Bohr and adopt the mesh spacings and $k$-point grids of: $0.15$ Bohr and no $k$-point for the molecule, $0.13$ Bohr and $14$ for the wire, and $0.15$ Bohr and $8\times8$ for the monolayer.
\subsection{Static polarizability}
\begin{figure}[!htbp]
    \centering
    \subfloat[4,5-Diaminophthalonitrile molecule \label{Fig:mol_pol}]{
        \includegraphics[width=0.49\textwidth,keepaspectratio=true]{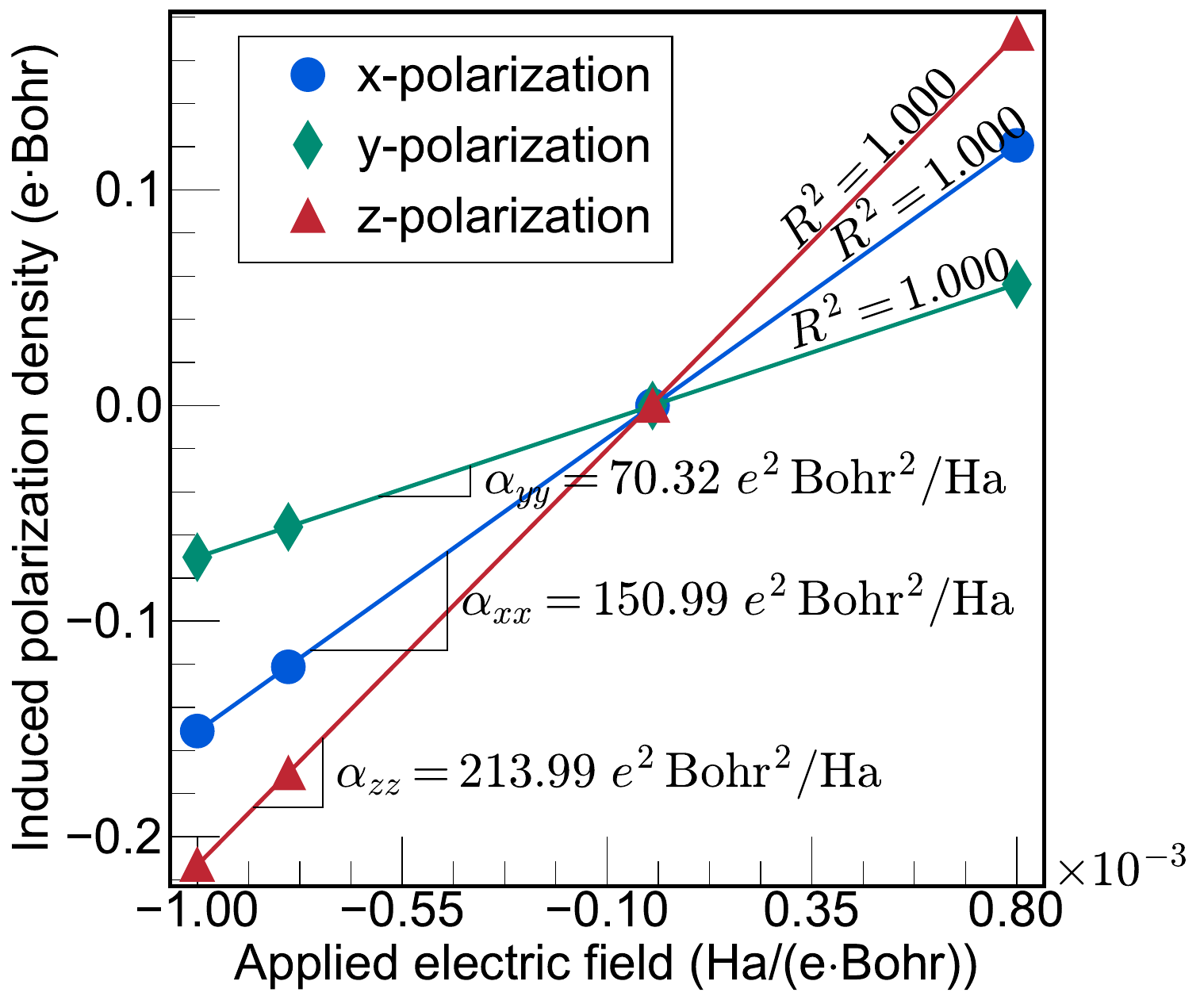}}
    \subfloat[Polycarbonitrile wire \label{Fig:wire_pol}]{
        \includegraphics[width=0.49\textwidth,keepaspectratio=true]{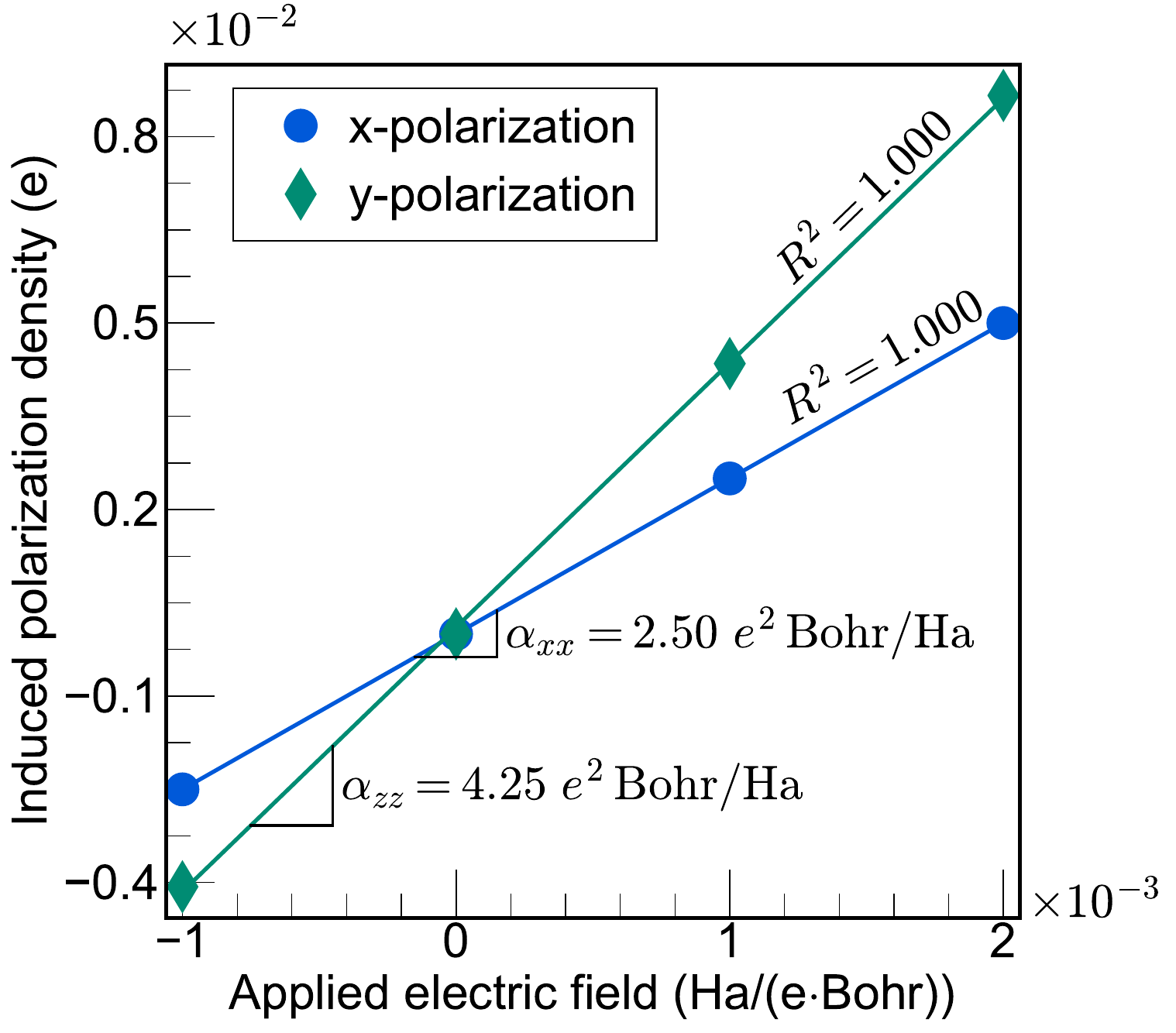}}\\
    \subfloat[Molybdenum sulfoselenide monolayer \label{Fig:surf_pol}]{
        \includegraphics[width=0.49\textwidth,keepaspectratio=true]{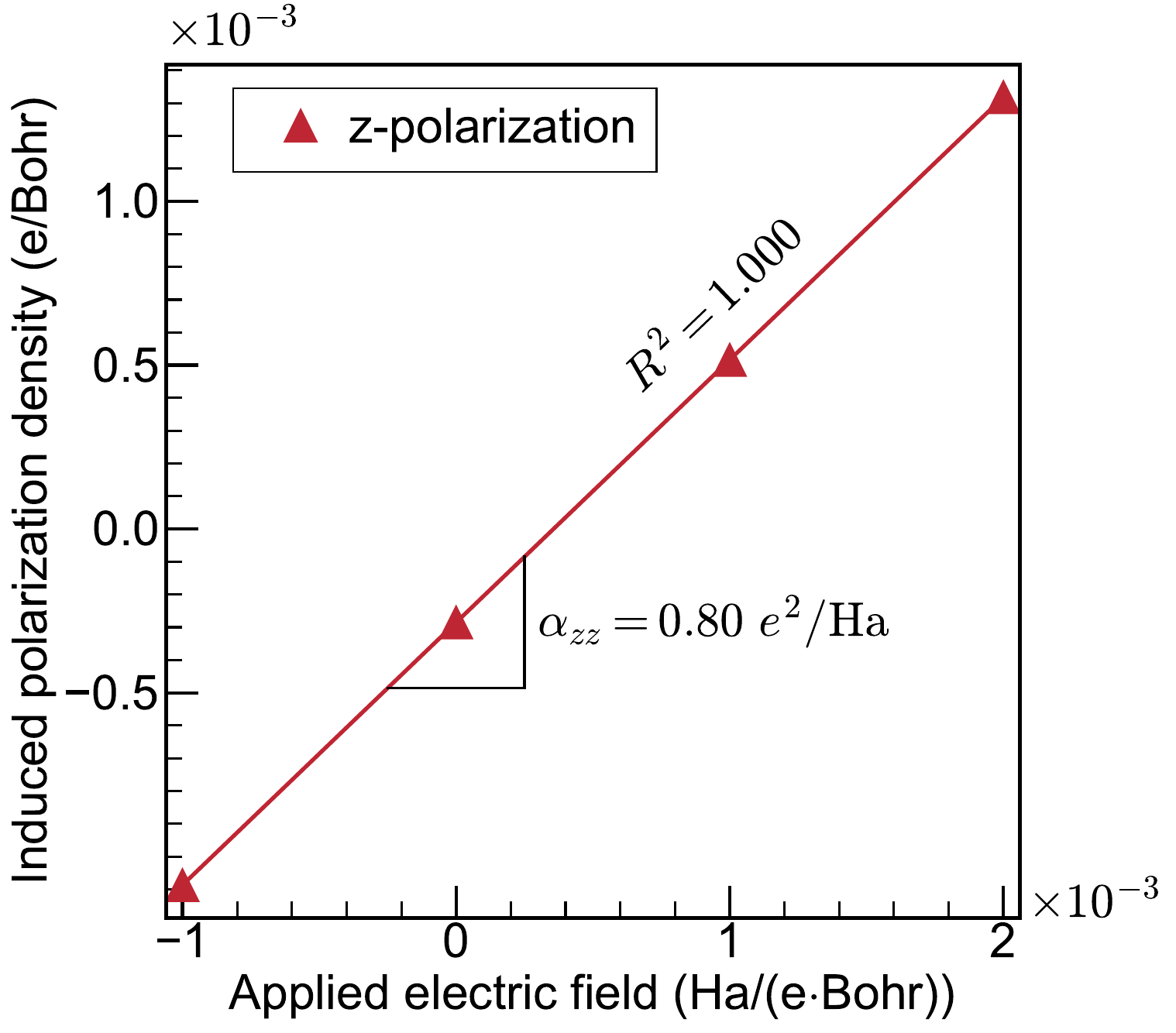}}
    \caption{Induced polarization density as a function of the applied electric field for (a) the 4,5-diaminophthalonitrile molecule, (b) the polycarbonitrile wire, and (c) the molybdenum sulfoselenide monolayer. Markers are computed values and solid lines are linear fits; the slopes give the polarizability components $\alpha_{ii}$, and the $R^2$ values indicate the linearity of the response.}
    \label{Fig:polarizability}
\end{figure}
The static polarizability $\alpha_{ij}$ of a material measures the linear response of its polarization density in the $i$th direction, $P_i$, to a uniform electric field applied to it in the $j$th direction, $E_j$, i.e.,
\begin{align}
 \alpha_{ij} = \frac{\partial P_i}{\partial E_j}\bigg|_{\bE=\mathbf{0}}.
 \end{align}
For each system, we compute all polarizability components $\alpha_{ij}$ with $i,j \in \cI_o$. We perform geometry relaxation after each application of the electric field to obtain both the electronic as well as ionic contribution to the polarizability. Fig.~\ref{Fig:polarizability} shows the induced polarization density as a function of the applied electric field. All systems exhibit a highly linear response, with $R^2 = 1.000$ in every case, confirming that the calculations lie well within the linear-response regime. For the molecule, all off-diagonal components are negligible and are therefore omitted; the diagonal components are $\alpha_{xx} = 150.99$, $\alpha_{yy} = 70.32$, and $\alpha_{zz} = 213.99~e^2\,\text{Bohr}^2/\text{Ha}$, the anisotropy reflecting the planar, elongated geometry of the molecule. For the wire, the off-diagonal component $\alpha_{xz}$ is likewise negligible, while the diagonal components are $\alpha_{xx} = 2.50$ and $\alpha_{zz} = 4.25~e^2\,\text{Bohr}/\text{Ha}$; their inequality reflects the anisotropic cross-section of the planar polycarbonitrile chain. For the monolayer, the out-of-plane component obtained is $\alpha_{zz} = 0.80$ e$^2$/Ha, which agrees closely with the value of $0.74$ e$^2$/Ha reported by Riis-Jensen et~al. \cite{riis2019classifying}. For each system, we also computed the clamped-ion polarizabilities. Relative to the relaxed-ion values reported above, these differ by $\sim \!3 \%$ (in $\alpha_{xx}$ and $\alpha_{zz}$) for the molecule and  by $\sim \!4 \%$ (in $\alpha_{zz}$) for the wire, and are essentially unchanged for the monolayer, indicating that the ionic contributions are small compared to the electronic ones.
\subsection{Piezoelectric coefficient}
\begin{figure}[!htbp]
    \centering
    \subfloat[Polycarbonitrile wire \label{Fig:wire_piezo}]{
        \includegraphics[width=0.49\textwidth,keepaspectratio=true]{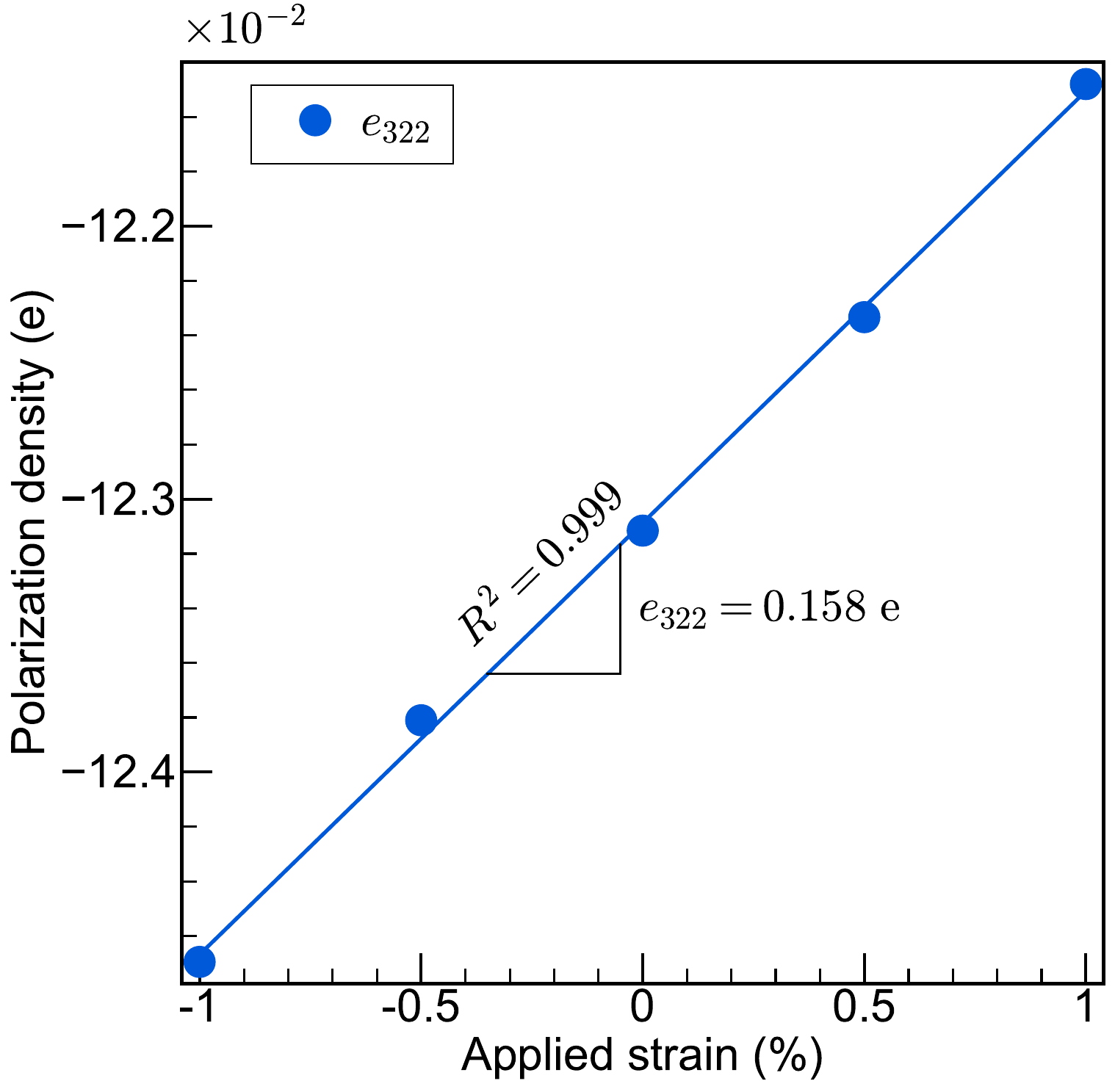}}
    \hfill
    \subfloat[Molybdenum sulfoselenide monolayer \label{Fig:surf_piezo}]{
        \includegraphics[width=0.49\textwidth,keepaspectratio=true]{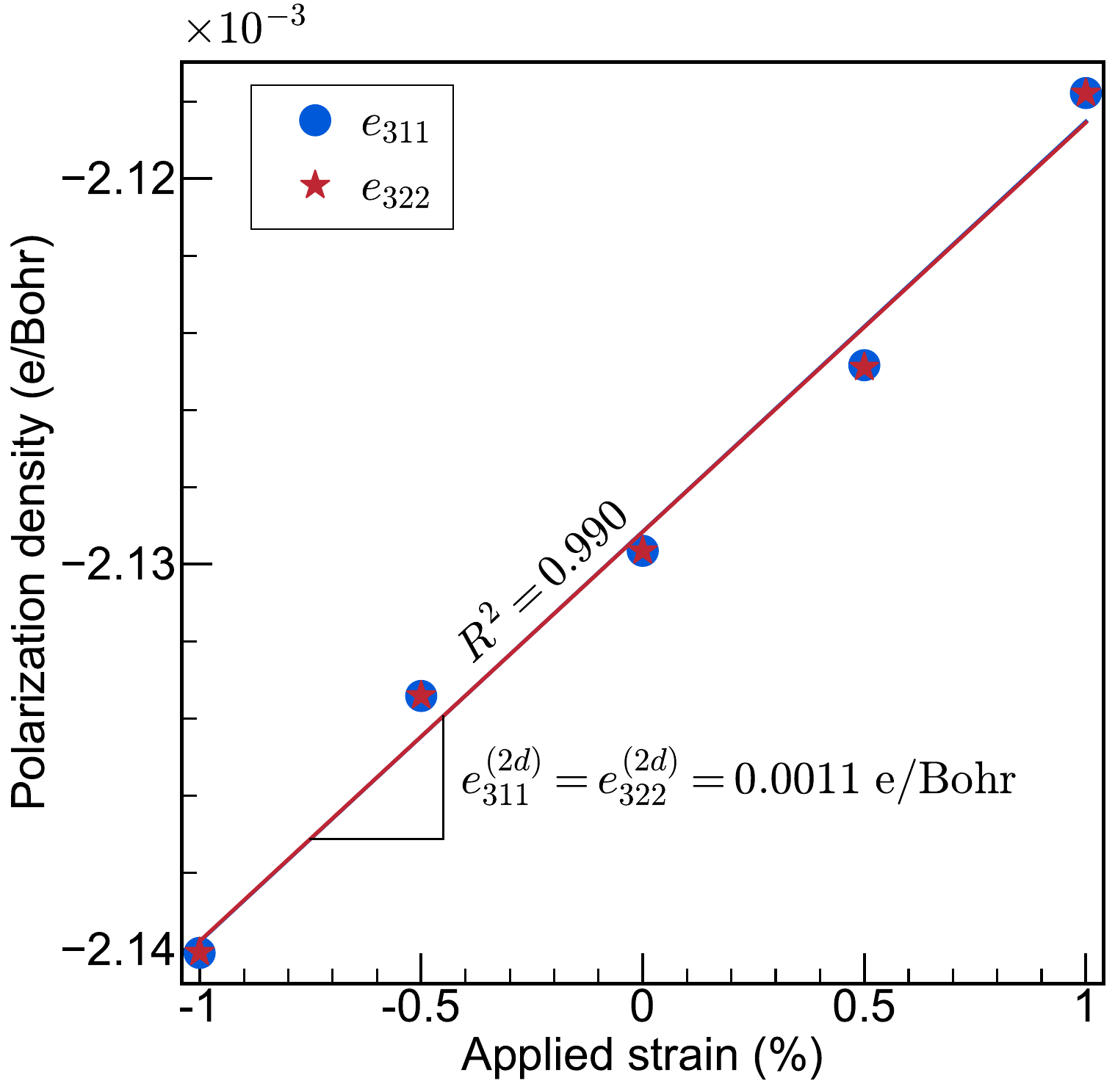}}
    \caption{Polarization density as a function of applied strain for (a) the polycarbonitrile wire and (b) the molybdenum sulfoselenide monolayer. Markers are computed values and solid lines are linear fits; the slopes give the piezoelectric coefficients $e_{3jj}$, with $R^2$ indicating the linearity of the response.}
    \label{Fig:piezoelectric}
\end{figure}

The piezoelectric coefficient $e_{ijk}$ of a material measures the linear response of its polarization density in the $i$th direction, $P_i$, to a mechanical strain $\eta_{jk}$ with $j,k \in \cI_p$, i.e.,
\begin{align}
 e_{ijk} = \frac{\partial P_i}{\partial \eta_{jk}}\bigg|_{\boldsymbol{\eta}=\mathbf{0}}.
 \end{align}
We now evaluate the piezoelectric coefficients of the wire and monolayer by applying uniform axial and shear strains and measuring the induced polarization density in their open directions after ionic relaxation. In all cases, the strain is varied over $\pm 1\%$ and the piezoelectric coefficients are extracted from a linear fit of the polarization density against the applied strain, as shown in Fig.~\ref{Fig:piezoelectric}. For the wire, we obtain $e_{322}$ as $0.158~e$ (Fig.~\ref{Fig:wire_piezo}), while $e_{122}$ is negligible and therefore omitted. Similarly, for the monolayer, we obtain the coefficients due to axial strains as $e_{311} = e_{322} = 0.0012$ e/Bohr (Fig.~\ref{Fig:surf_piezo}), while the $e_{312}$ component due to shear strain is negligible and therefore omitted. The two values due to axial strains are equal owing to the in-plane symmetry in the monolayer and both agree closely with the value of $0.0011$ e/Bohr reported in the literature \cite{dong2017large}. The near-unity values of $R^2$ confirm a well-defined, linear piezoelectric response within the strain range considered.
\section{Concluding remarks} \label{Sec:conclusion}
In this work, we developed an electrostatic formulation in real-space density functional theory that provides a systematic and unified treatment of the open-boundary electrostatics of isolated and partially periodic systems, including in the presence of an applied uniform electric field along the open directions. Specifically, we formulated an electrostatic energy functional whose stationarity yields the Poisson equation for the electrostatic potential---arising from the electronic and ionic charge densities as well as the applied uniform electric field---subject to periodic and Dirichlet boundary conditions along the periodic and open directions, respectively. The analytical expressions for the Dirichlet values arising from the charge density of the system were derived using a Green's function approach. On this basis, we developed the real-space formulation of Kohn--Sham DFT, deriving the electronic ground-state equations along with the expressions for the ground-state energy, atomic forces, and stress tensor. These were implemented within the large-scale parallel real-space electronic structure code SPARC \cite{xu2021sparc,zhang2024sparc}. Through representative examples, we verified the accuracy and efficiency of the formulation and its implementation. The computed quantities converged exponentially with respect to the vacuum size and were in excellent agreement with those obtained from established plane-wave codes, while requiring significantly less vacuum at comparable accuracy; the stresses agreed to within $0.1\%$ with those obtained from numerical derivatives of the energy. Finally, we applied the framework to compute static polarizabilities and piezoelectric coefficients, obtaining very good agreement with values reported in the literature.

The present formulation has several notable features: the electrostatic potential and/or its gradient is not presumed to vanish on the open boundaries; it is local and variational, and applies across dimensionalities within a single framework; the boundary values are prescribed for the total electrostatic potential rather than the Hartree potential alone; an applied uniform electric field along the open directions is incorporated; and the stress tensor is derived. Furthermore, in removing the excess vacuum padding and corrective terms hitherto necessary for isolated and partially periodic systems, the formulation reduces the computational domain to that consistent with the decay of the electron density. This is of particular relevance to low-dimensional materials with intrinsic polarization, polar surfaces and interfaces, and systems under applied electric fields. The availability of the stress tensor further enables cell relaxation and equation-of-state calculations for such systems.
 
As part of our future research, we plan to extend this electrostatic formulation to isolated and partially periodic systems carrying a net charge. In addition, the development of an analogous formulation for systems with cyclic and/or helical symmetry will enable the accurate and efficient study of electromechanical couplings \cite{codony2021transversal,kumar2021flexoelectricity} arising from bending \cite{kumar2020bending} and twisting \cite{bhardwaj2021torsional}. Finally, incorporating the first-order variations of the electrostatic quantities within the formalism of real-space density functional perturbation theory (DFPT) \cite{sharma2023calculation,sharma2026cyclic} would allow an accurate treatment of linear response properties of low-dimensional systems, and forms a promising avenue for future work.

\section*{Acknowledgements}
R.K. and A.S. gratefully acknowledge the ANRF early career research grant (Grant No. ANRF/ECRG/2024/002362) from Department of Science and Technology India. R.K. and A.S. also acknowledge the National Supercomputing Mission (NSM) for providing computing resources of ‘PARAM Ganga’ at the Indian Institute of Technology Roorkee, which is implemented by C-DAC and supported by the Ministry of Electronics and Information Technology (MeitY) and Department of Science and Technology (DST), Government of India. D.C. acknowledges the support of the Spanish Ministry of Universities through the project PID2023-152533OB-I00, funded by MICIU/AEI/10.13039/501100011033 and FEDER, and the Margarita Salas fellowship (European Union-NextGenerationEU)

\section*{Data Availability}
The data that support the findings of this article are openly available at \url{https://github.com/rajatkr544/Supporting_data}.

\appendix
\section{Effect of the nonzero Fourier components} \label{Sec:AppendixA}
\begin{figure}[!htbp]
    \centering
    \subfloat[Polycarbonitrile wire \label{Fig:longwire_vacuum}]{
        \includegraphics[width=0.49\textwidth,keepaspectratio=true]{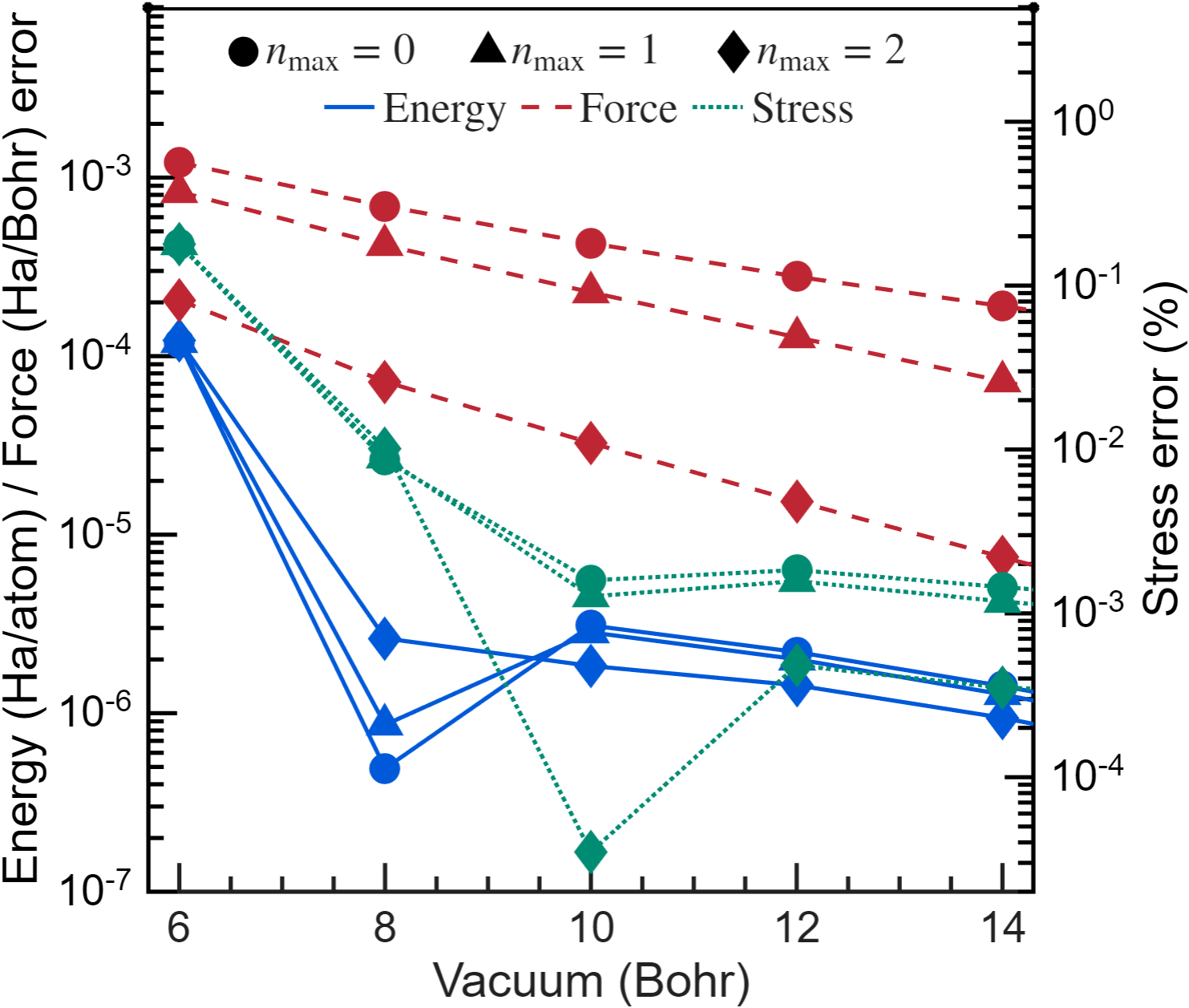}
    }
    \subfloat[Molybdenum sulfoselenide monolayer\label{Fig:largesurface_vacuum}]{
        \includegraphics[width=0.49\textwidth,keepaspectratio=true]{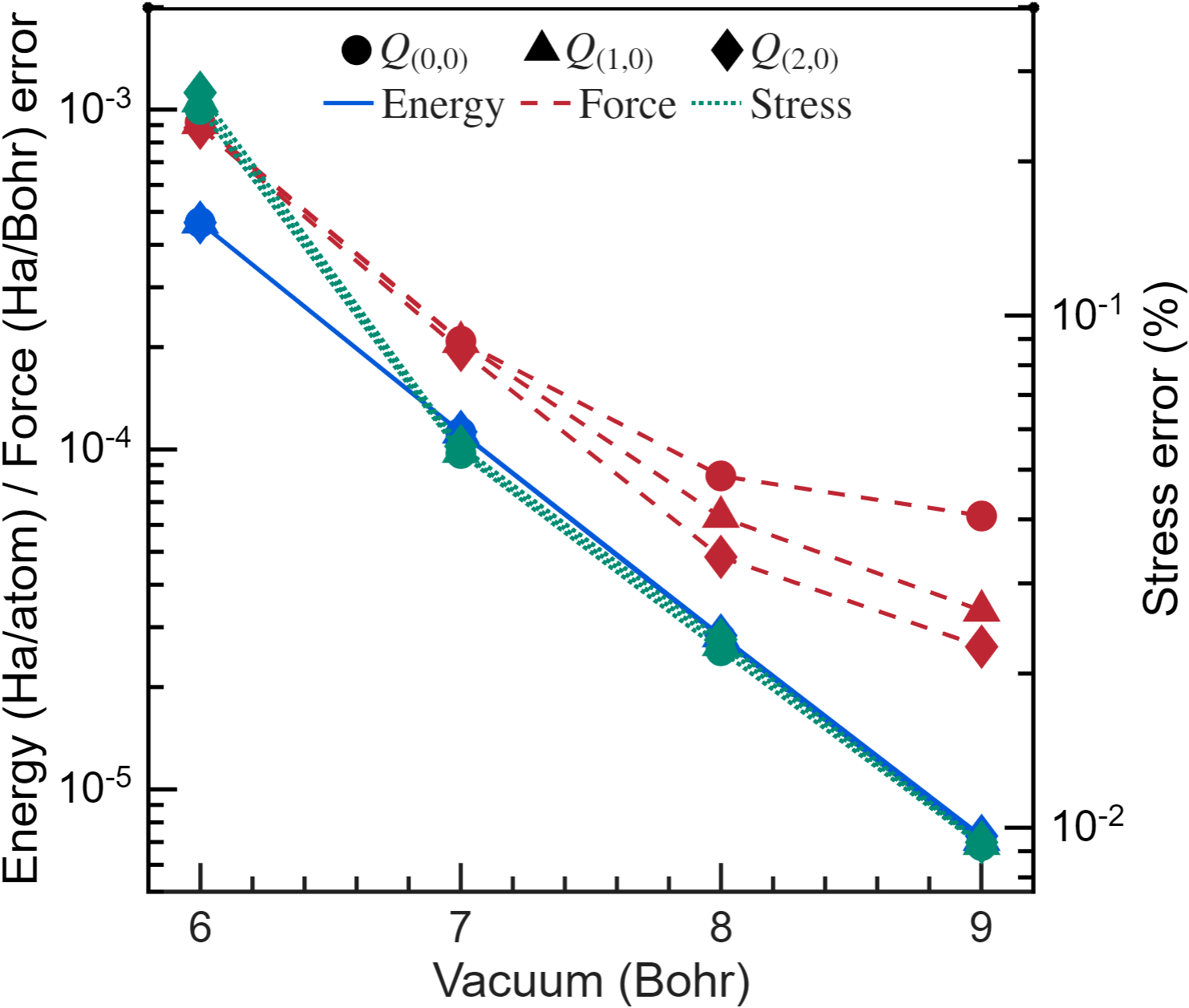}
    }
    \caption{Convergence of the energy (solid blue), forces (dashed red), and stresses (dotted green) with respect to vacuum size for (a) a long polycarbonitrile wire, and (b) a large molybdenum sulfoselenide monolayer, both with randomly perturbed atomic positions.} \label{Fig:highorder_vacuum}
\end{figure}

Here we examine the effect of the nonzero Fourier components, present in the boundary conditions of 1D and 2D periodic systems, on the convergence of the energy, atomic forces, and stresses. For this purpose, we choose a polycarbonitrile wire with $L_y = 100.24$ Bohr and a molybdenum sulfoselenide monolayer with $L_x = 105$ Bohr and $L_y = 6.14$ Bohr. In both systems, the atoms within the unit cell are randomly perturbed to obtain representatives of partially periodic systems with large supercells. An electric field of $5\times10^{-3}$ Ha/(e$\cdot$Bohr) is applied along the $y$- and $z$-directions of the wire, and $10^{-2}$ Ha/(e$\cdot$Bohr) along the $z$-direction of the monolayer. We perform $\Gamma$-point calculations with a mesh spacing of $0.2$ Bohr for both systems. For the wire, $m_\text{max}=3$ is chosen in all its simulations. Fig.~\ref{Fig:highorder_vacuum} shows the convergence of the energy, forces, and stresses with respect to vacuum size at different truncation parameters for both the wire and monolayer. It is evident from Fig.~\ref{Fig:longwire_vacuum} that including terms with $n_\text{max} > 0$ improves the convergence of all quantities with vacuum size for the wire, with the improvement most pronounced for the forces and stresses. For the monolayer (Fig.~\ref{Fig:largesurface_vacuum}), the effect of the $Q_{mn}$ terms with $(m,n) \neq (0,0)$ is clearly evident for the forces but less pronounced for the energy and stresses.

\bibliographystyle{elsarticle-num}

\bibliography{Electrostatics_DFT}

\begin{thebibliography}{10}
\expandafter\ifx\csname url\endcsname\relax
  \def\url#1{\texttt{#1}}\fi
\expandafter\ifx\csname urlprefix\endcsname\relax\def\urlprefix{URL }\fi
\expandafter\ifx\csname href\endcsname\relax
  \def\href#1#2{#2} \def\path#1{#1}\fi

\bibitem{hohenberg1964inhomogeneous}
P.~Hohenberg, W.~Kohn, Inhomogeneous electron gas, Phys. Rev. 136~(3B) (1964)
  B864.

\bibitem{kohn1965self}
W.~Kohn, L.~J. Sham, Self-consistent equations including exchange and
  correlation effects, Phys. Rev. 140~(4A) (1965) A1133.

\bibitem{burke2012dft}
K.~Burke, Perspective on density functional theory, J. Chem. Phys. 136 (2012)
  150901.

\bibitem{becke2014dft}
A.~D. Becke, Perspective: Fifty years of density-functional theory in chemical
  physics, J. Chem. Phys. 140 (2014) 18A301.

\bibitem{pickett1989pseudopotential}
W.~E. Pickett, Pseudopotential methods in condensed matter applications,
  Comput. Phys. Rep. 9~(3) (1989) 115--197.

\bibitem{Martin2004}
R.~Martin, Electronic Structure: Basic theory and practical methods, Cambridge
  University Press, 2004.

\bibitem{kresse1996efficient}
G.~Kresse, J.~Furthm{\"u}ller, Efficient iterative schemes for ab initio
  total-energy calculations using a plane-wave basis set, Phys. Rev. B 54~(16)
  (1996) 11169.

\bibitem{clark2005first}
S.~J. Clark, M.~D. Segall, C.~J. Pickard, P.~J. Hasnip, M.~J. Probert,
  K.~Refson, M.~C. Payne, First principles methods using castep, Z.
  Kristallogr. (2005) 567--570.

\bibitem{gonze2002first}
X.~Gonze, J.-M. Beuken, R.~Caracas, F.~Detraux, M.~Fuchs, G.-M. Rignanese,
  L.~Sindic, M.~Verstraete, G.~Zerah, F.~Jollet, et~al., First-principles
  computation of material properties: the abinit software project, Comput.
  Mater. Sci. 25~(3) (2002) 478--492.

\bibitem{giannozzi2009quantum}
P.~Giannozzi, S.~Baroni, N.~Bonini, M.~Calandra, R.~Car, C.~Cavazzoni,
  D.~Ceresoli, G.~L. Chiarotti, M.~Cococcioni, I.~Dabo, et~al., Quantum
  espresso: a modular and open-source software project for quantum simulations
  of materials, J. Phys. Condens. Matter 21~(39) (2009) 395502.

\bibitem{ismail2000new}
S.~Ismail-Beigi, T.~Arias, New algebraic formulation of density functional
  calculation, Comput. Phys. Commun. 128~(1-2) (2000) 1--45.

\bibitem{gygi2008architecture}
F.~Gygi, Architecture of qbox: A scalable first-principles molecular dynamics
  code, IBM J. Res. Dev. 52~(1.2) (2008) 137--144.

\bibitem{valiev2010nwchem}
M.~Valiev, E.~J. Bylaska, N.~Govind, K.~Kowalski, T.~P. Straatsma, H.~J.~J.
  Van~Dam, D.~Wang, J.~Nieplocha, E.~Apr{\`a}, T.~L. Windus, et~al., Nwchem: A
  comprehensive and scalable open-source solution for large scale molecular
  simulations, Comput. Phys. Commun. 181~(9) (2010) 1477--1489.

\bibitem{ewald1921berechnung}
P.~P. Ewald, Die berechnung optischer und elektrostatischer gitterpotentiale,
  Ann. Phys. (Leipzig) 369~(3) (1921) 253--287.

\bibitem{ihm1979momentum}
J.~Ihm, A.~Zunger, M.~L. Cohen, Momentum-space formalism for the total energy
  of solids, J. Phys. C: Solid State Phys. 12~(21) (1979) 4409--4422.

\bibitem{makov1995periodic}
G.~Makov, M.~C. Payne, Periodic boundary conditions in ab initio calculations,
  Phys. Rev. B. 51~(7) (1995) 4014.

\bibitem{kunc1983external}
K.~Kunc, R.~Resta, External fields in the self-consistent theory of electronic
  states: a new method for direct evaluation of macroscopic and microscopic
  dielectric response, Phys. Rev. Lett. 51~(8) (1983) 686.

\bibitem{neugebauer1992adsorbate}
J.~Neugebauer, M.~Scheffler, Adsorbate-substrate and adsorbate-adsorbate
  interactions of na and k adlayers on al (111), Phys. Rev. B 46~(24) (1992)
  16067.

\bibitem{bengtsson1999dipole}
L.~Bengtsson, Dipole correction for surface supercell calculations, Phys. Rev.
  B. 59~(19) (1999) 12301.

\bibitem{meyer2001ab}
B.~Meyer, D.~Vanderbilt, Ab initio study of {BaTiO$_3$} and {PbTiO$_3$}
  surfaces in external electric fields, Phys. Rev. B 63~(20) (2001) 205426.

\bibitem{rozzi2006exact}
C.~A. Rozzi, D.~Varsano, A.~Marini, E.~K. Gross, A.~Rubio, Exact coulomb cutoff
  technique for supercell calculations, Phys. Rev. B. 73~(20) (2006) 205119.

\bibitem{ismail2006truncation}
S.~Ismail-Beigi, Truncation of periodic image interactions for confined
  systems, Phys. Rev. B 73~(23) (2006) 233103.

\bibitem{dabo2008electrostatics}
I.~Dabo, B.~Kozinsky, N.~E. Singh-Miller, N.~Marzari, Electrostatics in
  periodic boundary conditions and real-space corrections, Phys. Rev. B 77~(11)
  (2008) 115139.

\bibitem{martyna1999reciprocal}
G.~J. Martyna, M.~E. Tuckerman, A reciprocal space based method for treating
  long range interactions in ab initio and force-field-based calculations in
  clusters, J. Chem. Phys. 110~(6) (1999) 2810.

\bibitem{barnett1993born}
R.~N. Barnett, U.~Landman, Born-oppenheimer molecular-dynamics simulations of
  finite systems: Structure and dynamics of (h 2 o) 2, Phys. Rev. B 48~(4)
  (1993) 2081.

\bibitem{sohier2017density}
T.~Sohier, M.~Calandra, F.~Mauri, Density functional perturbation theory for
  gated two-dimensional heterostructures: Theoretical developments and
  application to flexural phonons in graphene, Phys. Rev. B 96~(7) (2017)
  075448.

\bibitem{rivano2024density}
N.~Rivano, N.~Marzari, T.~Sohier, Density functional perturbation theory for
  one-dimensional systems: Implementation and relevance for phonons and
  electron-phonon interactions, Phys. Rev. B 109~(24) (2024) 245426.

\bibitem{freysoldt2008screening}
C.~Freysoldt, P.~Eggert, P.~Rinke, A.~Schindlmayr, M.~Scheffler, Screening in
  two dimensions: Gw calculations for surfaces and thin films using the
  repeated-slab approach, Phys. Rev. B: Condens. Matter Mater. Phys. 77~(23)
  (2008) 235428.

\bibitem{becke1989basis}
A.~D. Becke, Basis-set-free density-functional quantum chemistry, Int. J.
  Quantum Chem. 36~(S23) (1989) 599--609.

\bibitem{white1989finite}
S.~R. White, J.~W. Wilkins, M.~P. Teter, Finite-element method for electronic
  structure, Phys. Rev. B 39~(9) (1989) 5819.

\bibitem{chel1994fdpp}
J.~R. Chelikowsky, N.~Troullier, Y.~Saad, Finite-difference-pseudopotential
  method: Electronic structure calculations without a basis, Phys. Rev. Lett.
  72~(8) (1994) 1240--1243.

\bibitem{seitsonen1995real}
A.~P. Seitsonen, M.~J. Puska, R.~M. Nieminen, Real-space electronic-structure
  calculations: Combination of the finite-difference and conjugate-gradient
  methods, Phys. Rev. B 51~(20) (1995) 14057.

\bibitem{tsuchida1995electronic}
E.~Tsuchida, M.~Tsukada, Electronic-structure calculations based on the
  finite-element method, Phys. Rev. B 52~(8) (1995) 5573.

\bibitem{briggs1996real}
E.~Briggs, D.~Sullivan, J.~Bernholc, Real-space multigrid-based approach to
  large-scale electronic structure calculations, Phys. Rev. B 54~(20) (1996)
  14362.

\bibitem{fattebert1999finite}
J.-L. Fattebert, Finite difference schemes and block rayleigh quotient
  iteration for electronic structure calculations on composite grids, J.
  Comput. Phys. 149~(1) (1999) 75--94.

\bibitem{arias1999wav}
T.~A. Arias, Multiresolution analysis of electronic structure: semicardinal and
  wavelet bases, Rev. Mod. Phys. 71~(1) (1999) 267--311.

\bibitem{shimojo2001linear}
F.~Shimojo, R.~K. Kalia, A.~Nakano, P.~Vashishta, Linear-scaling
  density-functional-theory calculations of electronic structure based on
  real-space grids: design, analysis, and scalability test of parallel
  algorithms, Comput. Phys. Commun. 140~(3) (2001) 303--314.

\bibitem{skylaris2005introducing}
C.-K. Skylaris, P.~D. Haynes, A.~A. Mostofi, M.~C. Payne, Introducing onetep:
  Linear-scaling density functional simulations on parallel computers, J. Chem.
  Phys. 122~(8) (2005) 084119.

\bibitem{pask2005femeth}
J.~E. Pask, P.~A. Sterne, Finite element methods in ab initio electronic
  structure calculations, Model. Simul. Mater. Sci. Eng. 13 (2005) R71--R96.

\bibitem{bowler2006recent}
D.~Bowler, R.~Choudhury, M.~Gillan, T.~Miyazaki, Recent progress with
  large-scale ab initio calculations: the conquest code, Phys. Status Solidi B.
  243~(5) (2006) 989--1000.

\bibitem{castro2006octopus}
A.~Castro, H.~Appel, M.~Oliveira, C.~A. Rozzi, X.~Andrade, F.~Lorenzen, M.~A.
  Marques, E.~Gross, A.~Rubio, Octopus: a tool for the application of
  time-dependent density functional theory, Phys. Status Solidi B. 243~(11)
  (2006) 2465--2488.

\bibitem{genovese2008daubechies}
L.~Genovese, A.~Neelov, S.~Goedecker, T.~Deutsch, S.~A. Ghasemi, A.~Willand,
  D.~Caliste, O.~Zilberberg, M.~Rayson, A.~Bergman, et~al., Daubechies wavelets
  as a basis set for density functional pseudopotential calculations, The J.
  Chem. Phys. 129~(1) (2008) 014109.

\bibitem{iwata2010massively}
J.-I. Iwata, D.~Takahashi, A.~Oshiyama, T.~Boku, K.~Shiraishi, S.~Okada,
  K.~Yabana, A massively-parallel electronic-structure calculations based on
  real-space density functional theory, J. Comput. Phys. 229~(6) (2010)
  2339--2363.

\bibitem{suryanarayana2010non}
P.~Suryanarayana, V.~Gavini, T.~Blesgen, K.~Bhattacharya, M.~Ortiz,
  Non-periodic finite-element formulation of kohn--sham density functional
  theory, J. Mech. Phys. Solids 58~(2) (2010) 256--280.

\bibitem{suryanarayana2011mesh}
P.~Suryanarayana, K.~Bhattacharya, M.~Ortiz, A mesh-free convex approximation
  scheme for kohn--sham density functional theory, J. Comput. Phys. 230~(13)
  (2011) 5226--5238.

\bibitem{lin2012adaptive}
L.~Lin, J.~Lu, L.~Ying, et~al., Adaptive local basis set for kohn--sham density
  functional theory in a discontinuous galerkin framework {I}: Total energy
  calculation, J. Comput. Phys. 231~(4) (2012) 2140--2154.

\bibitem{Ghosh2017cluster}
S.~Ghosh, P.~Suryanarayana, Sparc: Accurate and efficient finite-difference
  formulation and parallel implementation of density functional theory:
  Isolated clusters, Comput. Phys. Commun. 212 (2017) 189--204.

\bibitem{Ghosh2017extended}
S.~Ghosh, P.~Suryanarayana, Sparc: Accurate and efficient finite-difference
  formulation and parallel implementation of density functional theory:
  Extended systems, Comput. Phys. Commun. 216 (2017) 109--125.

\bibitem{xu2018discrete}
Q.~Xu, P.~Suryanarayana, J.~E. Pask, Discrete discontinuous basis projection
  method for large-scale electronic structure calculations, J. Chem. Phys.
  149~(9) (2018) 094104.

\bibitem{motamarri2020dft}
P.~Motamarri, S.~Das, S.~Rudraraju, K.~Ghosh, D.~Davydov, V.~Gavini, Dft-fe--a
  massively parallel adaptive finite-element code for large-scale density
  functional theory calculations, Comput. Phys. Commun. 246 (2020) 106853.

\bibitem{beck2000rsmeth}
T.~L. Beck, Real-space mesh techniques in density-functional theory, Rev. Mod.
  Phys. 72~(4) (2000) 1041--1080.

\bibitem{saad2010esmeth}
Y.~Saad, J.~R. Chelikowsky, S.~M. Shontz, Numerical methods for electronic
  structure calculations of materials, SIAM Rev. 52~(1) (2010) 3--54.

\bibitem{pratapa2015spectral}
P.~P. Pratapa, P.~Suryanarayana, J.~E. Pask, Spectral quadrature method for
  accurate o (n) electronic structure calculations of metals and insulators,
  Comput. Phys. Commun. (2015).

\bibitem{suryanarayana2018sqdft}
P.~Suryanarayana, P.~P. Pratapa, A.~Sharma, J.~E. Pask, Sqdft: Spectral
  quadrature method for large-scale parallel o (n) kohn--sham calculations at
  high temperature, Comput. Phys. Commun. 224 (2018) 288--298.

\bibitem{gavini2023roadmap}
V.~Gavini, S.~Baroni, V.~Blum, D.~R. Bowler, A.~Buccheri, J.~R. Chelikowsky,
  S.~Das, W.~Dawson, P.~Delugas, M.~Dogan, et~al., Roadmap on electronic
  structure codes in the exascale era, Model. Simul. Mater. Sci. Eng. 31~(6)
  (2023) 063301.

\bibitem{banerjee2016cyclic}
A.~S. Banerjee, P.~Suryanarayana, Cyclic density functional theory: A route to
  the first principles simulation of bending in nanostructures, J. Mech. Phys.
  Solids 96 (2016) 605--631.

\bibitem{ghosh2019symmetry}
S.~Ghosh, A.~S. Banerjee, P.~Suryanarayana, Symmetry-adapted real-space density
  functional theory for cylindrical geometries: Application to large group-iv
  nanotubes, Phys. Rev. B 100~(12) (2019) 125143.

\bibitem{sharma2021real}
A.~Sharma, P.~Suryanarayana, Real-space density functional theory adapted to
  cyclic and helical symmetry: Application to torsional deformation of carbon
  nanotubes, Phys. Rev. B. 103~(3) (2021) 035101.

\bibitem{Gavini2007}
V.~Gavini, J.~Knap, K.~Bhattacharya, M.~Ortiz, Non-periodic finite-element
  formulation of orbital-free density functional theory, J. Mech. Phys. Solids
  55~(4) (2007) 669 -- 696.

\bibitem{alemany2004real}
M.~Alemany, M.~Jain, L.~Kronik, J.~R. Chelikowsky, Real-space pseudopotential
  method for computing the electronic properties of periodic systems, Phys.
  Rev. B 69~(7) (2004) 075101.

\bibitem{hirose2005first}
K.~Hirose, T.~Ono, Y.~Fujimoto, S.~Tsukamoto, First-principles calculations in
  real-space formalism (2005).

\bibitem{han2008real}
J.~Han, M.~L. Tiago, T.-L. Chan, J.~R. Chelikowsky, Real space method for the
  electronic structure of one-dimensional periodic systems, J. Chem. Phys.
  129~(14) (2008) 144109.

\bibitem{natan2008real}
A.~Natan, A.~Benjamini, D.~Naveh, L.~Kronik, M.~L. Tiago, S.~P. Beckman, J.~R.
  Chelikowsky, Real-space pseudopotential method for first principles
  calculations of general periodic and partially periodic systems, Phys. Rev.
  B—Condensed Matter and Materials Physics 78~(7) (2008) 075109.

\bibitem{ramakrishnan2025real}
K.~Ramakrishnan, G.~Sai~Gautam, P.~Motamarri, Real-space methods for ab initio
  modeling of surfaces and interfaces under external potential bias, J. Chem.
  Theory Comput. 21~(14) (2025) 7087--7101.

\bibitem{xu2021sparc}
Q.~Xu, A.~Sharma, B.~Comer, H.~Huang, E.~Chow, A.~J. Medford, J.~E. Pask,
  P.~Suryanarayana, Sparc: Simulation package for ab-initio real-space
  calculations, SoftwareX 15 (2021) 100709.

\bibitem{zhang2024sparc}
B.~Zhang, X.~Jing, Q.~Xu, S.~Kumar, A.~Sharma, L.~Erlandson, S.~J. Sahoo,
  E.~Chow, A.~J. Medford, J.~E. Pask, et~al., Sparc v2. 0.0: Spin-orbit
  coupling, dispersion interactions, and advanced exchange--correlation
  functionals, Softw. Impacts 20 (2024) 100649.

\bibitem{lennard1928cohesion}
J.~Lennard-Jones, B.~M. Dent, Cohesion at a crystal surface, Trans. Faraday
  Soc. 24 (1928) 92--108.

\bibitem{abramowitz1964handbook}
M.~Abramowitz, I.~A. Stegun, Handbook of Mathematical Functions with Formulas,
  Graphs, and Mathematical Tables, National Bureau of Standards, Washington,
  D.C., 1964.

\bibitem{bhowmik2026bulk}
S.~Bhowmik, A.~J. Medford, P.~Suryanarayana, Bulk boundary condition for
  surface calculations in density functional theory, arXiv preprint
  arXiv:2607.07894 (2026).

\bibitem{mermin1965thermal}
N.~D. Mermin, Thermal properties of the inhomogeneous electron gas, Phys. Rev.
  137~(5A) (1965) A1441.

\bibitem{kleinman1982efficacious}
L.~Kleinman, D.~Bylander, Efficacious form for model pseudopotentials, Phys.
  Rev. Lett. 48~(20) (1982) 1425.

\bibitem{harris1985simplified}
J.~Harris, Simplified method for calculating the energy of weakly interacting
  fragments, Phys. Rev. B 31~(4) (1985) 1770.

\bibitem{foulkes1989tight}
W.~M.~C. Foulkes, R.~Haydock, Tight-binding models and density-functional
  theory, Phys. Rev. B 39~(17) (1989) 12520.

\bibitem{feynman1939forces}
R.~P. Feynman, Forces in molecules, Phys. Rev. 56~(4) (1939) 340.

\bibitem{sharma2018calculation}
A.~Sharma, P.~Suryanarayana, On the calculation of the stress tensor in
  real-space kohn-sham density functional theory, J. Chem. Phys. 149~(19)
  (2018) 194104.

\bibitem{sharma2023gpu}
A.~Sharma, A.~Metere, P.~Suryanarayana, L.~Erlandson, E.~Chow, J.~E. Pask, Gpu
  acceleration of local and semilocal density functional calculations in the
  sparc electronic structure code, J. Chem. Phys. 158~(20) (2023).

\bibitem{jing2025gpu}
X.~Jing, A.~Sharma, J.~E. Pask, P.~Suryanarayana, Gpu acceleration of hybrid
  functional calculations in the sparc electronic structure code, J. Chem.
  Phys. 162~(18) (2025).

\bibitem{suryanarayana2014augmented}
P.~Suryanarayana, D.~Phanish, Augmented lagrangian formulation of orbital-free
  density functional theory, J. Comput. Phys. 275 (2014) 524--538.

\bibitem{zhou2006self}
Y.~Zhou, Y.~Saad, M.~L. Tiago, J.~R. Chelikowsky, Self-consistent-field
  calculations using chebyshev-filtered subspace iteration, J. Comput. Phys.
  219~(1) (2006) 172--184.

\bibitem{zhou2006parallel}
Y.~Zhou, Y.~Saad, M.~L. Tiago, J.~R. Chelikowsky, Parallel
  self-consistent-field calculations via chebyshev-filtered subspace
  acceleration, Phys. Rev. E 74~(6) (2006) 066704.

\bibitem{pratapa2015restarted}
P.~P. Pratapa, P.~Suryanarayana, Restarted pulay mixing for efficient and
  robust acceleration of fixed-point iterations, Chem. Phys. Lett. 635 (2015)
  69--74.

\bibitem{Banerjee2016PeriodicPulay}
A.~S. Banerjee, P.~Suryanarayana, J.~E. Pask, Periodic pulay method for robust
  and efficient convergence acceleration of self-consistent field iterations,
  Chem. Phys. Lett. 647 (2016) 31 -- 35.

\bibitem{realspaceprecond}
S.~Kumar, Q.~Xu, P.~Suryanarayana, On preconditioning the self-consistent field
  iteration in real-space density functional theory, Chem. Phys. Lett. 739
  (2020) 136983.

\bibitem{suryanarayana2019alternating}
P.~Suryanarayana, P.~P. Pratapa, J.~E. Pask, Alternating anderson--richardson
  method: An efficient alternative to preconditioned krylov methods for large,
  sparse linear systems, Comput. Phys. Commun. 234 (2019) 278--285.

\bibitem{pratapa2016anderson}
P.~P. Pratapa, P.~Suryanarayana, J.~E. Pask, Anderson acceleration of the
  jacobi iterative method: An efficient alternative to krylov methods for
  large, sparse linear systems, J. Comput. Phys. 306 (2016) 43--54.

\bibitem{press2007numerical}
W.~H. Press, Numerical recipes 3rd edition: The art of scientific computing,
  Cambridge university press, 2007.

\bibitem{monkhorst1976special}
H.~J. Monkhorst, J.~D. Pack, Special points for brillouin-zone integrations,
  Phys. Rev. B 13~(12) (1976) 5188.

\bibitem{hamann2013optimized}
D.~Hamann, Optimized norm-conserving vanderbilt pseudopotentials, Phys. Rev. B
  88~(8) (2013) 085117.

\bibitem{spms}
M.~F. Shojaei, J.~E. Pask, A.~J. Medford, P.~Suryanarayana, Soft and
  transferable pseudopotentials from multi-objective optimization, Comput.
  Phys. Commun. 283 (2023) 108594.

\bibitem{perdew1996generalized}
J.~P. Perdew, K.~Burke, M.~Ernzerhof, Generalized gradient approximation made
  simple, Phys. Rev. Lett. 77~(18) (1996) 3865.

\bibitem{riis2019classifying}
A.~C. Riis-Jensen, T.~Deilmann, T.~Olsen, K.~S. Thygesen, Classifying the
  electronic and optical properties of janus monolayers, ACS nano 13~(11)
  (2019) 13354--13364.

\bibitem{dong2017large}
L.~Dong, J.~Lou, V.~B. Shenoy, Large in-plane and vertical piezoelectricity in
  janus transition metal dichalchogenides, ACS nano 11~(8) (2017) 8242--8248.

\bibitem{codony2021transversal}
D.~Codony, I.~Arias, P.~Suryanarayana, Transversal flexoelectric coefficient
  for nanostructures at finite deformations from first principles, Phys. Rev.
  Mater. 5~(3) (2021) L030801.

\bibitem{kumar2021flexoelectricity}
S.~Kumar, D.~Codony, I.~Arias, P.~Suryanarayana, Flexoelectricity in atomic
  monolayers from first principles, Nanoscale 13~(3) (2021) 1600--1607.

\bibitem{kumar2020bending}
S.~Kumar, P.~Suryanarayana, Bending moduli for forty-four select atomic
  monolayers from first principles, Nanotechnology 31~(43) (2020) 43LT01.

\bibitem{bhardwaj2021torsional}
A.~Bhardwaj, A.~Sharma, P.~Suryanarayana, Torsional strain engineering of
  transition metal dichalcogenide nanotubes: an ab initio study, Nanotechnology
  32~(47) (2021) 47LT01.

\bibitem{sharma2023calculation}
A.~Sharma, P.~Suryanarayana, Calculation of phonons in real-space density
  functional theory, Phys. Rev. E 108~(4) (2023) 045302.

\bibitem{sharma2026cyclic}
A.~Sharma, P.~Suryanarayana, Cyclic-and helical-symmetry-adapted phonon
  formalism within density functional perturbation theory, Phys. Rev. B
  113~(20) (2026) 205116.

\end{thebibliography}


\end{document}